%% file: manuscript.tex
\documentclass[preprint]{elsarticle}\usepackage[]{graphicx}\usepackage[]{color}
\makeatletter
\def\maxwidth{ %
  \ifdim\Gin@nat@width>\linewidth
    \linewidth
  \else
    \Gin@nat@width
  \fi
}
\makeatother

\definecolor{fgcolor}{rgb}{0.345, 0.345, 0.345}
\newcommand{\hlnum}[1]{\textcolor[rgb]{0.686,0.059,0.569}{#1}}%
\newcommand{\hlstr}[1]{\textcolor[rgb]{0.192,0.494,0.8}{#1}}%
\newcommand{\hlcom}[1]{\textcolor[rgb]{0.678,0.584,0.686}{\textit{#1}}}%
\newcommand{\hlopt}[1]{\textcolor[rgb]{0,0,0}{#1}}%
\newcommand{\hlstd}[1]{\textcolor[rgb]{0.345,0.345,0.345}{#1}}%
\newcommand{\hlkwa}[1]{\textcolor[rgb]{0.161,0.373,0.58}{\textbf{#1}}}%
\newcommand{\hlkwb}[1]{\textcolor[rgb]{0.69,0.353,0.396}{#1}}%
\newcommand{\hlkwc}[1]{\textcolor[rgb]{0.333,0.667,0.333}{#1}}%
\newcommand{\hlkwd}[1]{\textcolor[rgb]{0.737,0.353,0.396}{\textbf{#1}}}%

\usepackage{framed}
\makeatletter
\newenvironment{kframe}{%
 \def\at@end@of@kframe{}%
 \ifinner\ifhmode%
  \def\at@end@of@kframe{\end{minipage}}%
  \begin{minipage}{\columnwidth}%
 \fi\fi%
 \def\FrameCommand##1{\hskip\@totalleftmargin \hskip-\fboxsep
 \colorbox{shadecolor}{##1}\hskip-\fboxsep
     \hskip-\linewidth \hskip-\@totalleftmargin \hskip\columnwidth}%
 \MakeFramed {\advance\hsize-\width
   \@totalleftmargin\z@ \linewidth\hsize
   \@setminipage}}%
 {\par\unskip\endMakeFramed%
 \at@end@of@kframe}
\makeatother

\definecolor{shadecolor}{rgb}{.97, .97, .97}
\definecolor{messagecolor}{rgb}{0, 0, 0}
\definecolor{warningcolor}{rgb}{1, 0, 1}
\definecolor{errorcolor}{rgb}{1, 0, 0}
\newenvironment{knitrout}{}{} 

\usepackage{alltt}         
\usepackage[margin=2.5cm]{geometry}
\usepackage{lineno}

\modulolinenumbers[5]

\journal{NeuroImage}

\usepackage{etoolbox}
\patchcmd{\pprintMaketitle}
 {\ifvoid\absbox\else\unvbox\absbox\par\vskip10pt\fi}
 {\ifvoid\absbox\else\clearpage\unvbox\absbox\par\vskip30pt\fi}
 {}{}
\patchcmd{\pprintMaketitle}
 {\hrule\vskip12pt}
 {}
 {}{}
\patchcmd{\pprintMaketitle}
 {\hrule\vskip12pt}
 {}
 {}{}
\appto{\pprintMaketitle}{\clearpage}

\usepackage{pdfpages} 
\usepackage{cprotect}
\usepackage{enumitem}
\usepackage{graphicx}
\usepackage{cprotect}
\usepackage{makecell}
\usepackage[hidelinks]{hyperref}
\usepackage{natbib}
\usepackage{booktabs} 
\usepackage{amsmath,amsfonts,amssymb}   
\usepackage{tabularx}
\usepackage{mdframed}

\setcitestyle{round}

\biboptions{authoryear}

\usepackage{newfloat}
\DeclareFloatingEnvironment[
    fileext=tb,
    listname={Boxes},
    name=Box,
    placement=tbhp
]{textbox}

\makeatletter
\def\ps@pprintTitle{%
   \let\@oddhead\@empty
   \let\@evenhead\@empty
   \let\@oddfoot\@empty
   \let\@evenfoot\@oddfoot
}
\makeatother
\IfFileExists{upquote.sty}{\usepackage{upquote}}{}
\begin{document}

\title{A recipe for accurate estimation of lifespan brain trajectories, distinguishing longitudinal and cohort effects}

\begin{frontmatter}

\author[1]{{\O}ystein S{\o}rensen\corref{corrauth}}
\cortext[corrauth]{Corresponding author: {\O}ystein S{\o}rensen, Department of Psychology, University of Oslo, Pb. 1094 Blindern, 0317 Oslo, Norway.}
\ead{oystein.sorensen@psykologi.uio.no}

\author[1,2]{Kristine B. Walhovd}
\author[1,2]{Anders M. Fjell}

\address[1]{Center for Lifespan Changes in Brain and Cognition, Department of Psychology, University of Oslo, Norway}
\address[2]{Department of Radiology and Nuclear Medicine, Oslo University Hospital, Norway}

\begin{abstract}
We address the problem of estimating how different parts of the brain develop and change throughout the lifespan, and how these trajectories are affected by genetic and environmental factors. Estimation of these lifespan trajectories is statistically challenging, since their shapes are typically highly nonlinear, and although true change can only be quantified by longitudinal examinations, as follow-up intervals in neuroimaging studies typically cover less than 10 \% of the lifespan, use of cross-sectional information is necessary. Linear mixed models (LMMs) and structural equation models (SEMs) commonly used in longitudinal analysis rely on assumptions which are typically not met with lifespan data, in particular when the data consist of observations combined from multiple studies. While LMMs require a priori specification of a polynomial functional form, SEMs do not easily handle data with unstructured time intervals between measurements. Generalized additive mixed models (GAMMs) offer an attractive alternative, and in this paper we propose various ways of formulating GAMMs for estimation of lifespan trajectories of 12 brain regions, using a large longitudinal dataset and realistic simulation experiments. We show that GAMMs are able to more accurately fit lifespan trajectories, distinguish longitudinal and cross-sectional effects, and estimate effects of genetic and environmental exposures. Finally, we discuss and contrast questions related to lifespan research which strictly require repeated measures data and questions which can be answered with a single measurement per participant, and in the latter case, which simplifying assumptions that need to be made. The examples are accompanied with R code, providing a tutorial for researchers interested in using GAMMs.
\end{abstract}


\end{frontmatter}

\section{Introduction}

\begin{textbox}
\begin{mdframed}
\textbf{Cohort effect}: The effect of birth year (cohort) on the relationship between a set of explanatory variables and an outcome of interest.\\
\textbf{Cross-sectional effect}: The effect of age on an outcome of interest at a particular point in time, across participants with different birth dates.\\
\textbf{Longitudinal effect}: The effect of increasing age for participants belonging to a given birth cohort.\\
\textbf{Linear mixed models} (LMMs): Linear regression models used for data with hierarchical structure, e.g. longitudinal data with multiple measurements per individual.\\
\textbf{Fixed effects}: Regression parameters in mixed models which are common for all participants.\\
\textbf{Random effects}: Regression parameters in mixed models which are unique to each participant, used to model correlation between repeated measurements.\\
\textbf{Polynomial model}: Linear regression model which includes the first $n$ powers of an explanatory variable $x$ as distinct variables.\\
\textbf{Quadratic model}: A polynomial model containing the first two powers of an explanatory variable $x$, on the form $\beta_{0} + \beta_{1} x + \beta_{2} x^{2}$.\\
\textbf{Cubic model}: A polynomial model containing the first three powers of an explanatory variable $x$, on the form $\beta_{0} + \beta_{1} x + \beta_{2} x^{2} + \beta_{3} x^{3}$.\\
\textbf{Generalized additive model} (GAM): A linear regression model in which the outcome is modeled as an unknown smooth function of the explanatory variables.\\
\textbf{Generalized additive mixed model} (GAMM): An extension of GAMs to data with hierarchical structure, containing random effects.\\
\textbf{Smoothing parameter}: Parameter controlling the degree of nonlinearity (wiggliness) of a function estimated by a GAM/GAMM.\\
\textbf{Cubic regression splines}: A set of cubic polynomials, each of which is defined over a small part of the $x$-axis and is zero elsewhere.\\
\textbf{Thin-plate regression splines}: A set of functions, each of which represents a given nonlinear shape over the full $x$-axis.\\
\textbf{Smooth function}: A nonlinear function estimated by GAMs/GAMMs represented as a weighted sum of (e.g., cubic or thin-plate) regression splines.
\end{mdframed}
\caption{Key terms used in this paper, defined in the context of longitudinal data analysis.}
\label{box:definition}
\end{textbox}

\begin{figure}
\centering
\includegraphics[width=\columnwidth]{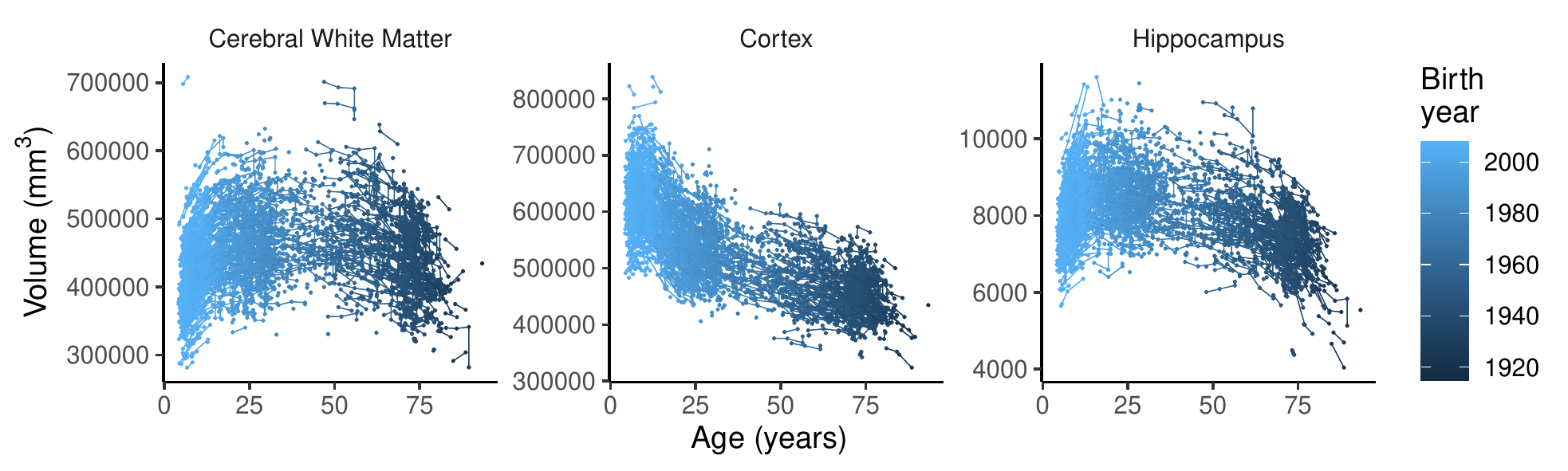}
\cprotect\caption{\textbf{Lifespan brain development is highly nonlinear}. Cerebral white matter, cortex and hippocampal volumes from 4,352 MRI scans of 2,017 participants in the LCBC data. The color scale indicates the birth cohort to which the participant belongs. Dots represent observations and lines connecting the dots indicate repeated observations of the same individual.}
\label{fig:spaghetti_plots}
\end{figure}

\begin{figure}
\centering
\includegraphics{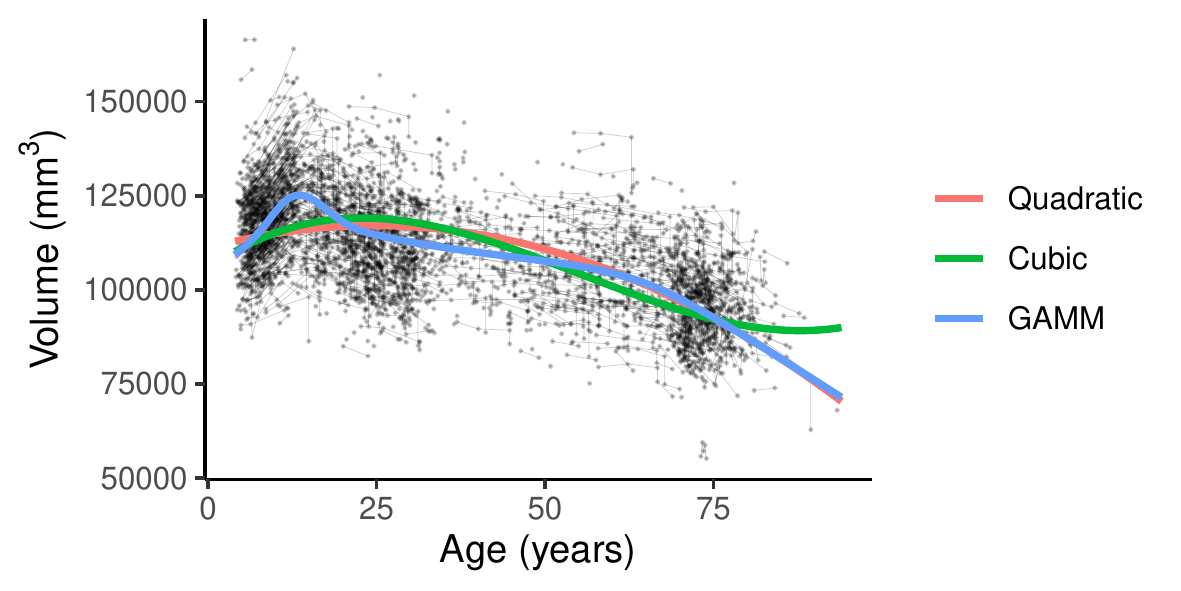}
\cprotect\caption{\textbf{Comparison of LMMs and GAMMs for lifespan data}. Comparison of LMMs with quadratic and cubic terms and a GAMM, fitted to lifespan cerebellum cortex volume. Black dots represent observations and black lines connecting the dots indicate repeated observations of the same individual.}
\label{fig:gamm_vs_poly}
\end{figure}

\begin{figure}
\centering
\includegraphics[width=\columnwidth]{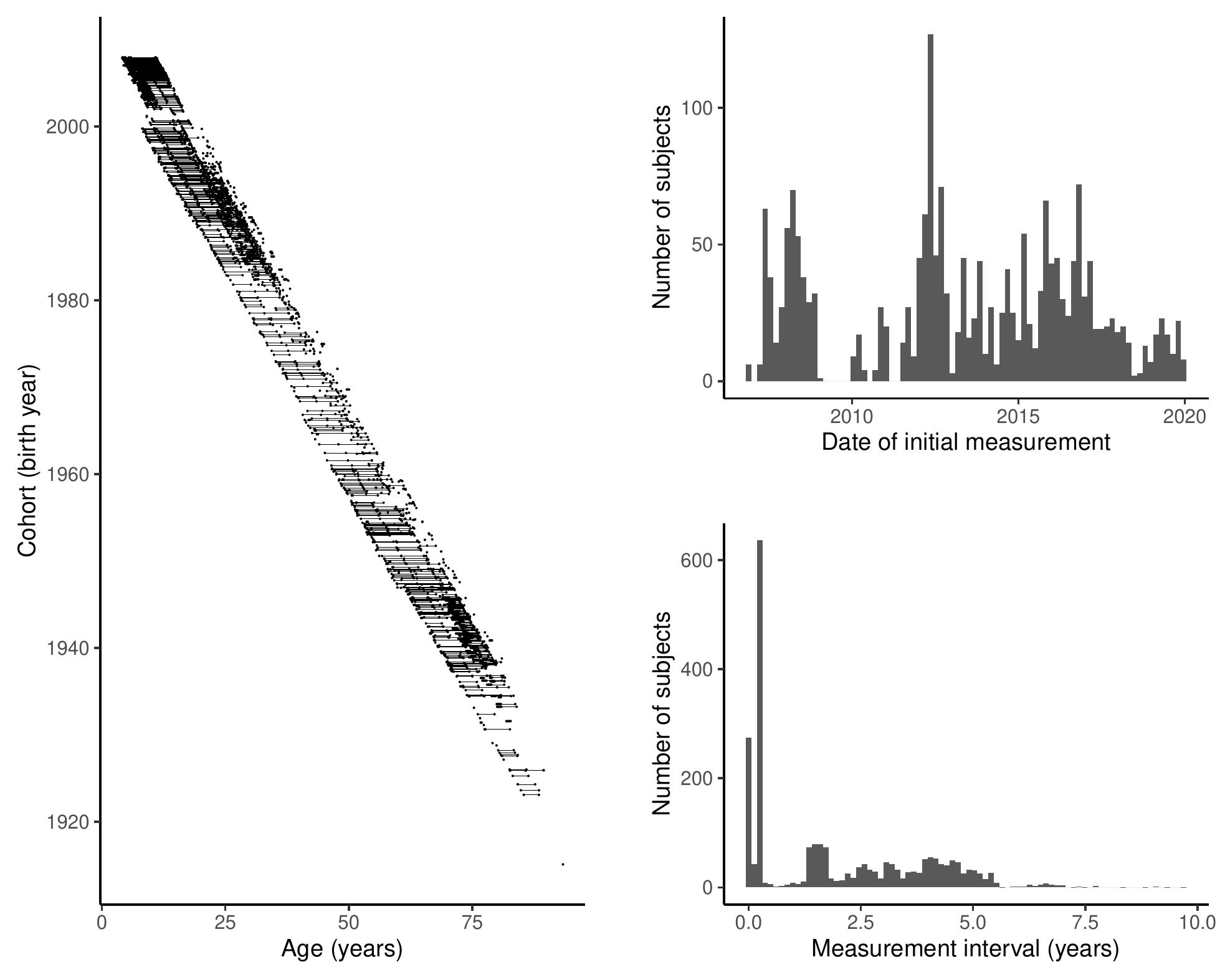}
\cprotect\caption{\textbf{Characteristics of lifespan data}. The plots show data from 4,352 MRI scans of 2,017 participants in the LCBC data. Left: Scatter plot of age and cohort. Connected dots show repeated measurements of the same participant. Top right: Histogram of date of initial measurement for the same participants. Bottom right: Histogram of time (in years) between measurements. The peak at zero corresponds to participants scanned twice on the same day, with different scanners, and the highest peak corresponds to participants with 10-11 weeks between measurements.}
\label{fig:date_dist}
\end{figure}

Large datasets with structural magnetic resonance images (MRIs) of participants whose ages span from early childhood to late adulthood provide ample opportunities to study lifespan brain trajectories. Important questions such data can contribute to answer include how brain structure is related to aging, how the aging effect is modified by genetics and environmental exposures, and at which age critical events like maximum volume or maximum rate of change occur. Lifespan brain trajectories are nonlinear and differ between regions, as illustrated in Figure \ref{fig:spaghetti_plots} for volumes of cerebral white matter, cortex, and hippocampus for 4,352 observations of 2,017 healthy participants from the Center for Lifespan Changes in Brain and Cognition longitudinal studies \citep{Fjell2017,Walhovd2016}, henceforth referred to as the LCBC data. Modeling the type of nonlinear effects shown in Figure \ref{fig:spaghetti_plots} using linear mixed models (LMMs) \citep{Laird1982} with polynomials typically leads to poor fits at least over parts of the lifespan, and is highly dependent on manual selection of terms \citep{Fjell2010,Sorensen2021}. Structural equation models (SEMs) may be better able to estimate nonlinear trajectories, e.g., with a latent basis model \citep{McArdle1987,Meredith1990}, but SEMs require that the time intervals between measurements for all participants take on a small set of unique values \citep{Newsom2015,Oud2000}, an assumption which may be hard to satisfy with lifespan data (see Figure \ref{fig:date_dist}). Generalized additive mixed models (GAMMs)\footnote{We will use the common abbreviation "GAMM", although strictly speaking only additive mixed models (AMMs) are used in this paper. If necessary, all models described can be straightforwardly generalized, e.g. to logistic or Poisson regression.} \citep{Lin1999} offer an attractive alternative, typically yielding good fit over the full lifespan in an automated and data-driven manner. This is illustrated in Figure \ref{fig:gamm_vs_poly}, comparing a GAMM to LMMs with quadratic and cubic polynomials for the effect of age on cerebellum cortex volume. See Box \ref{box:definition} for a definition of these and other key terms used in this paper. Similar to often researched structures like hippocampus and cerebral white matter, cerebellum cortex is characterized by a nonlinear age trajectory. In contrast to the GAMM, neither of the LMMs capture the steep increase seen in early childhood, and the cubic LMM predicts an increase in cerebellum cortex volume in old age, whereas the GAMM adequately captures the decline seen in the data. In addition, both the quadratic and cubic model estimate cerebellum cortex volume to reach its maximum at the age of around 25, while the GAMM instead estimates the maximum to occur around 14 years of age, and the latter seems to be in better agreement with the data. Figures \ref{fig:spaghetti_plots} and \ref{fig:gamm_vs_poly} and all subsequent plots were created in \verb!R! \citep{Rcore} with \verb!ggplot2! \citep{Wickham2016}.

The goal of this paper is to provide clear recommendations for optimal estimation of lifespan trajectories of brain development and aging. To this end, several aspects need consideration. First, as has been emphasized by a large number of authors, when analyzing data with repeated observations over time, care must be taken to distinguish within-individual and between-individual effects, which for the purpose of this paper are longitudinal and cross-sectional effects \citep{Curran2011,Hoffman2007,Hoffman2009,Morrell2009,Sliwinski2010,Thompson2011}. Individual change can only be assessed with repeated measurements, but how important are longitudinal data when the task is to estimate trajectories spanning many times the maximum follow-up interval realistically attainable in a neuroimaging study?  Large datasets combined from different studies, either conducted by the same group as for the LCBC data or by multiple groups participating in a data-sharing consortium like Lifebrain \citep{Walhovd2018} or a meta-analysis network like ENIGMA \citep{Bearden2017,Thompson2017}, present further challenges for longitudinal modeling as the number of measurements per participant and the time intervals between measurements are typically highly varying. All of these issues are illustrated for the LCBC data in Figure \ref{fig:date_dist}. While GAMMs flexibly handle data with varying follow-up intervals, the statistical literature on use of GAMMs for longitudinal analysis has almost exclusively focused on cases in which each participant has been followed over the full time range under consideration, from a common baseline \citep{Brumback1998,Durban2005,Edwards2005,Gu2005,Ke2001,Lambert2001,Sullivan2015}. There is hence a need for an understanding of how GAMMs should be optimally used in lifespan brain research, with short follow-up intervals and varying dates of initial measurement as shown in Figure \ref{fig:date_dist}.

The outline of this paper is as follows. In Section \ref{sec:methods} we introduce GAMMs formally and define three different candidate models for estimating lifespan brain trajectories. We also describe simulation experiments conducted in order to compare these GAMMs in a realistic setting for estimating 12 different brain regions. In Section \ref{sec:results} the simulation results are presented, and next we show two example applications demonstrating how GAMMs can be used for estimating lifespan brain trajectories and the effect of genetic variations on the trajectories. Accompanying \verb!R! code provides a tutorial for researchers interested in using GAMMs. In Section \ref{sec:discussion} we discuss the results taking into regard currently available longitudinal studies. We contrast questions that strictly require longitudinal data to questions that under some simplifying assumptions may be answered with a single measurement per participant. Finally, in Section \ref{sec:conclusion} we conclude by presenting recommendations for how to use GAMMs in lifespan brain research.

\section{Materials and methods}
\label{sec:methods}

\subsection{Longitudinal and cross-sectional effects}
\begin{figure}
\centering
\includegraphics[width=\columnwidth]{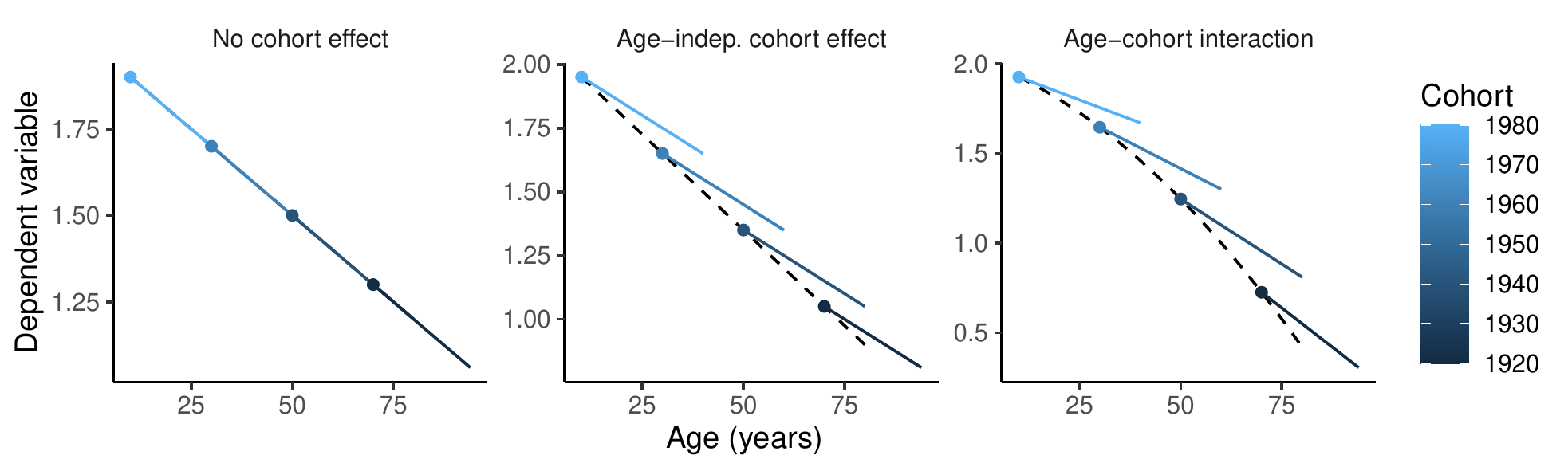}
\caption{\textbf{Cohort effects}. Illustration of the impact of cohort effects in a hypothetical dataset. Dashed lines show the cross-sectional age effect in 1990, colored dots show four cohorts of participants whose age in 1990 was 10, 30, 50, and 70 years, respectively, and the blue lines show longitudinal age effects for each cohort. In the left plot, there are no cohort effects, and hence longitudinal and cross-sectional effects coincide. In the center plot, the cohort effects are independent of age, and the longitudinal effects differ by an offset but the effect of aging is identical across cohorts, as seen by the parallel blue lines. In the right plot, the cohort effects depend on age, and in this case also the slope of the longitudinal effect varies between cohorts.}
\label{fig:cohort_effects_illustration}
\end{figure}

The effect of age on an outcome in a population can be completely explained by longitudinal and cohort effects, with the former representing the effect of aging for participants in a given birth cohort and the latter determining how the longitudinal effects differ between participants belonging to different birth cohorts \citep[Ch. 1.1]{Diggle2002}. Cross-sectional effects are the effects of age across cohorts when considered at a particular point in time, as illustrated in Figure \ref{fig:cohort_effects_illustration}. In the absence of cohort effects the cross-sectional and longitudinal effects are identical. Age-independent cohort effects result in different slopes for the longitudinal and cross-sectional effects, while age-cohort interactions lead to longitudinal effects whose slopes depend on age. Selective survival, by which life expectancy is correlated with the dependent variable, leads to population changes over time and hence are part of the longitudinal effects \citep{Baltes1968}. In contrast, with sampling bias, by which the probability of recruitment or the probability of dropout before the end of the study depends on the outcome variable, the sample is not representative of the population under study and biased estimates may result \citep{Molenberghs2009}.

\subsection{Generalized additive models}
\label{sec:GAMintro}

Generalized additive models (GAMs) \citep{Hastie1986} model the effect of a variable $x$ on an outcome $y$ with smooth functions $f(x)$, constructed as weighted sums of $K$ basis functions $b_{1}(x)$, $b_{2}(x)$, $\dots$, $b_{K}(x)$ with weights $\beta_{1}$, $\beta_{2}$, $\dots$, $\beta_{K}$, i.e., $f(x) = \sum_{k=1}^{K} \beta_{k} b_{k}(x)$. Commonly used basis functions are cubic regression splines and thin-plate regression splines \citep{Wood2003}, and the number of basis functions is typically chosen large enough to allow a wide range of nonlinear patterns to be estimated, while small enough to allow computational efficiency. For a GAM with a single smooth term, $y = f(x) + \epsilon$, the estimate given $n$ observations is computed by finding the values of $\beta_{1}, \dots, \beta_{K}$ minimizing the criterion
\begin{equation*}
\sum_{i=1}^{n} \left[y_{i} - \sum_{k=1}^{K} \beta_{k} b_{k}(x_{i}) \right]^{2} + \lambda \int_{a}^{b}\left[ \sum_{k=1}^{K} \beta_{k} b_{k}''(x)\right]^2 \text{d}x.
\end{equation*}
The first term is the least squares criterion using the basis functions as explanatory variables, and the second term represents the wiggliness of $f(x)$ as measured by its squared second derivative over some range $[a,b]$, typically the minimum and maximum values of $x$ in the sample. The smoothing parameter $\lambda$ controls the extent to which wiggliness is penalized, striking a balance between overfitting (too low $\lambda$, too wiggly $f(x)$) and underfitting (too high $\lambda$, too smooth $f(x)$). For data with repeated measurements, GAMs can be extended to GAMMs by the inclusion of random effects. A key insight allowing use of LMM software for efficient fitting of GAMMs is that the penalized smooth terms may be decomposed into a fixed effect part representing unpenalized linear functional forms with zero second derivative, and a random effect part representing penalized nonlinear functional forms with non-zero second derivative \citep{Lin1999,Wood2004,Wood2010}. The variance of the random effects is proportional to $1/\lambda$, and this allows the smoothing parameter to be estimated as a mixed model variance component.

\subsection{Generalized additive mixed models for longitudinal data}
\label{sec:GAMMlong}

In this section we present three different models for estimating lifespan brain trajectories.

Consider a dataset of $n$ participants indexed $i=1,\dots,n$, assume an outcome $y_{ij}$ has been measured $m_{i}$ times in participant $i$, with timepoints indexed by $j=1,\dots,m_{i}$, and let $a_{ij}$ denote the age of participant $i$ at her/his $j$th timepoint. The question of interest is how the outcome varies as a function of age, and this can be modeled with the GAMM
\begin{equation}
\label{eq:GAMM1}
y_{ij} = \beta_{0} + f(a_{ij}) + b_{0i} + \epsilon_{ij},
\end{equation}
where $f(a_{ij})$ is the effect of age, $\beta_{0}$ is the intercept, $b_{0i}$ is the random intercept for participant $i$, and $\epsilon_{ij}$ is a random noise term. Both $b_{0i}$ and $\epsilon_{ij}$ are assumed to be normally distributed, $b_{0i} \sim N(0, \sigma_{b}^{2})$ and $\epsilon_{ij} \sim N(0, \sigma^2)$, with $\sigma_{b}$ representing the between-participant standard deviation and $\sigma$ the within-participant residual standard deviation. We do not consider random slopes, due to the low number of repeated measurements in the typical applications considered in this paper, although this could be included with an additional term $b_{1i}a_{ij}$ in \eqref{eq:GAMM1}. With sufficient data, use of random slopes is recommended, as it relaxes the assumptions on the covariance structure of repeated measurements \citep[Ch. 19]{Fitzmaurice2011}. 

In the presence of cohort effects, the term $f(a_{ij})$ represents some weighted combination of cross-sectional and longitudinal effects, and hence cannot be interpreted as either. The typical method of correcting for this in LMMs is by splitting the age term into $a_{i1}$ representing age at first measurement, and $t_{ij}$ representing time since baseline \citep{Fitzmaurice2011,Zeger1992} (see \citet{Mehta2000} for an equivalent method in SEMs). Extending this to a GAMM yields
\begin{equation}
\label{eq:GAMM2}
y_{ij} = \beta_{0} + f(a_{i1}, t_{ij}) +  b_{0i} + \epsilon_{ij},
\end{equation}
where $f(a_{i1},t_{ij})$ is a smooth bivariate function of baseline age and time. Considering the plots in Figure \ref{fig:spaghetti_plots}, using a bivariate function seems necessary for estimating lifespan trajectories, as the direction of change clearly depend on baseline age. In model \eqref{eq:GAMM2} the longitudinal effect of aging $t$ from a baseline $a_{i1}$ is given by $f(a_{i1}, t)$ keeping $a_{i1}$ constant, while the cross-sectional effect of varying baseline age $a$ is given by $f(a, 0)$.

Model \eqref{eq:GAMM2} has some important limitations, however. First, an assumption in the LMMs motivating its definition is that all participants have identical baseline dates. Second, when participants are followed over a short period compared to the full lifespan, the values of $t_{ij}$ vary between zero and some maximum which is much lower than the maximum age, which might make estimation of nonlinear longitudinal effects challenging. We hence introduce an alternative GAMM modeling cohort effects by including birth date $c_{i}$,
\begin{equation}
\label{eq:GAMM3}
y_{ij} = \beta_{0} + f(a_{ij}) + \beta(a_{ij})c_{i}  + b_{0i} + \epsilon_{ij},
\end{equation}
in which $f(a_{ij})$ is defined as for \eqref{eq:GAMM1}, while $\beta(a_{ij})c_{i}$ is a varying-coefficient term \citep{Hastie1993} representing the main effect of cohort (birth date) as a function of age. The longitudinal effect of aging $a$ for a participant belonging to cohort $c_{i}$ is given by $f(a) + \beta(a)c_{i}$ keeping $c_{i}$ constant. The cross-sectional effect of age $a$ at date $d$ is given by $f(a) + \beta(a)c$, with $c = d-a$ representing the birth date of participants of age $a$ at date $d$, and hence both $a$ and $c$ are varying in this case. Model \eqref{eq:GAMM3} is not identified if all participants have identical measurement dates, since then birth date and age are perfectly collinear, i.e., $c_{i} = d_{j} - a_{ij}$ where $d_{j}$ is the common date of the $j$th timepoint. However, as illustrated in Figure \ref{fig:date_dist} (right), both the dates of initial measurements and the times between measurements may be highly varying in lifespan data, and this variability helps identifying the estimates of model \eqref{eq:GAMM3}.

The effect of a time-invariant variable $x_{i}$ on the age trajectory can be estimated by adding an interaction term to models \eqref{eq:GAMM1}-\eqref{eq:GAMM3}. If $x_{i}$ is a continuous variable, the interaction may be a varying-coefficient term of the form introduced in model \eqref{eq:GAMM3}. For \eqref{eq:GAMM1} and \eqref{eq:GAMM3} it would be $\beta_{x}(a_{ij})x_{i}$, and for \eqref{eq:GAMM2} it would be $\beta_{x}(a_{i}, t_{ij})x_{i}$, where $\beta_{x}(\cdot)$ in both cases is a smooth function representing the effect of $x_{i}$ as a function of either age or baseline age and time. On the other hand, if $x_{i}$ is a categorical variable with $L$ unique values, it can be encoded as $L-1$ dummy variables $x_{2i},\dots, x_{Li}$, with a varying-coefficient term $\beta_{l}(a_{ij})x_{i}$ associated with the $l$th level for models \eqref{eq:GAMM1} and \eqref{eq:GAMM3}, and similarly $\beta_{l}(a_{i}, t_{ij})x_{i}$ for model \eqref{eq:GAMM2}. Each varying-coefficient term now represents how the trajectory for the $l$th level differs from the trajectory for the baseline level. An example is given in Section \ref{sec:interactions}.

\subsection{Simulation experiments}

\begin{figure}
\centering
\includegraphics{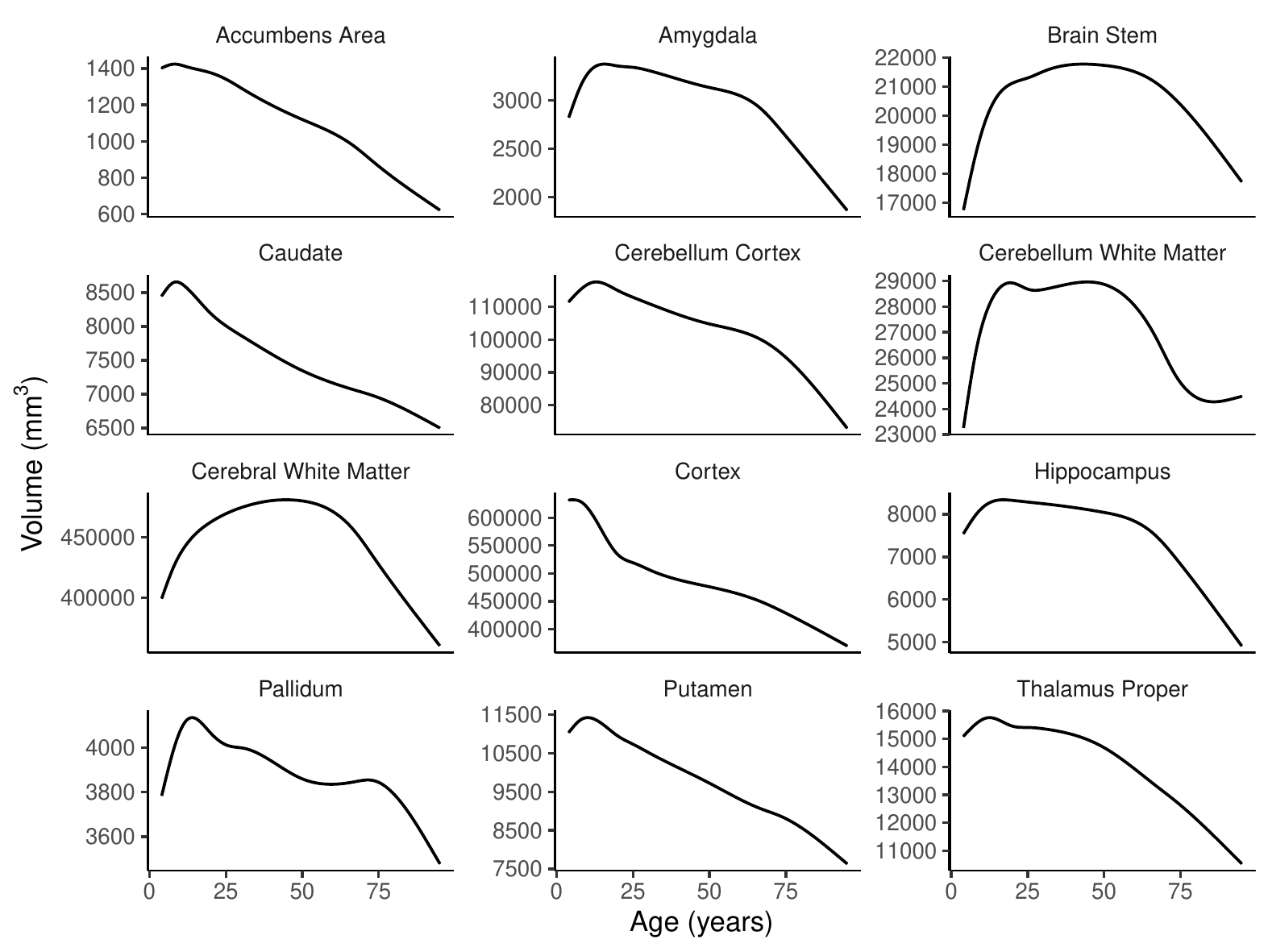}
\caption{\textbf{Lifespan curves}. Characteristic curves of 12 brain regions, estimated from the LCBC data and used in simulation experiments.}
\label{fig:characteristic_curves}
\end{figure}

\begin{figure}
\centering
\includegraphics[width=\columnwidth]{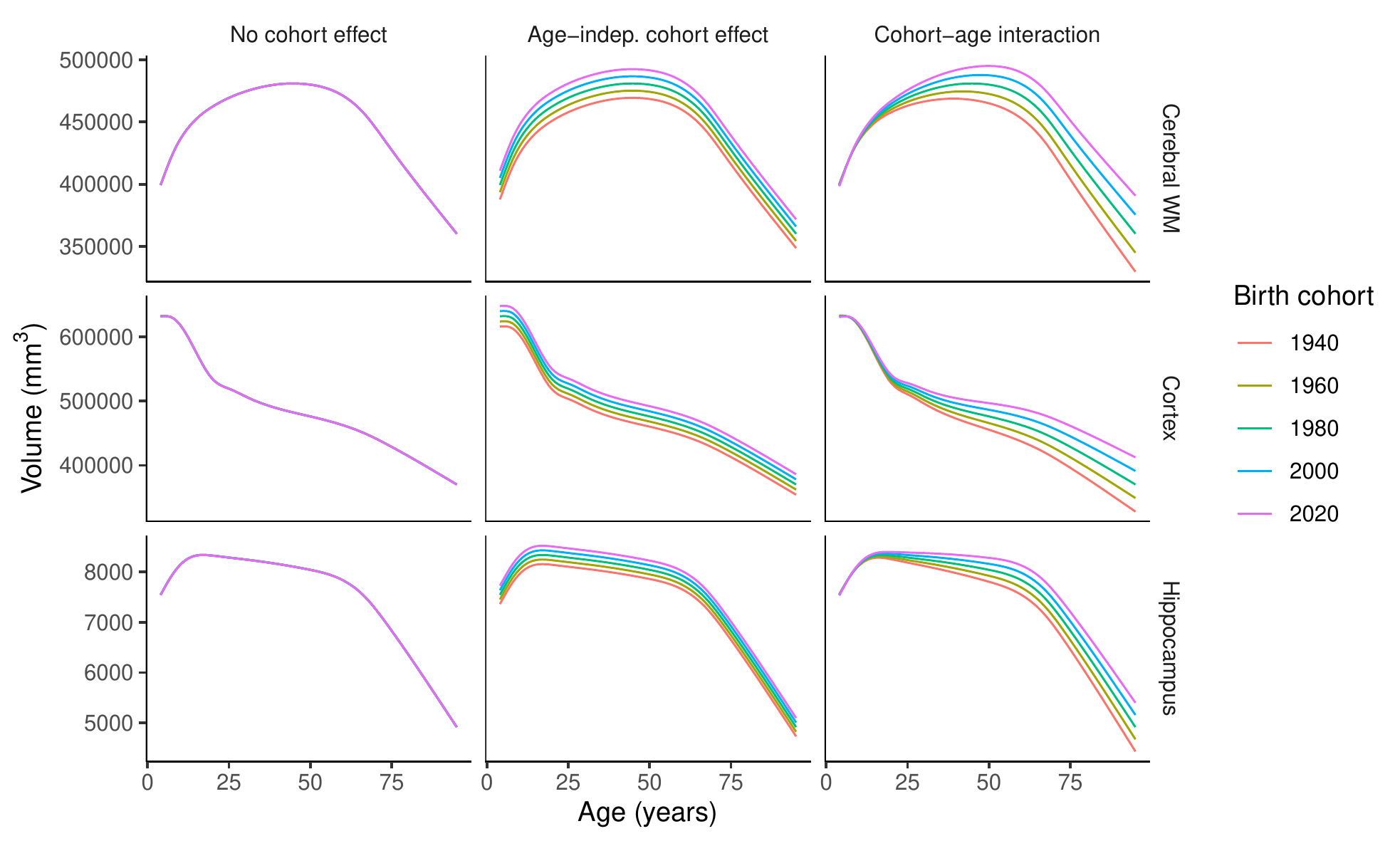}
\caption{\textbf{Simulated cohort effects}. Cohort effects used in simulation studies, illustrated for cerebral white matter, cortex, and hippocampus. (Cerebral WM = Cerebral White Matter)}
\label{fig:cohort_effects_main}
\end{figure}

In order to compare the GAMMs \eqref{eq:GAMM1}-\eqref{eq:GAMM3}, characteristic lifespan curves were estimated for 12 brain regions with the LCBC data, using GAMMs on the form \eqref{eq:GAMM1}, with additional covariates sex, scanner, and total intracranial volume (ICV). Volumes were estimated with FreeSurfer 6.0 \citep{Dale1999,Fischl2002,Reuter2012}, and detailed sample characteristics are presented in the Supplementary Section S1. The curves, shown in Figure \ref{fig:characteristic_curves}, were used as ground truths from which measurements were sampled. For each region, three cases were considered: no cohort effects, age-independent cohort effects, and age-cohort interactions. In the latter two cases, cohort effects were added to the characteristic curves as illustrated in Figure \ref{fig:cohort_effects_main} for cerebral white matter, cortex, and hippocampus, and in Supplementary Section S2.2 for the remaining regions. Data were generated with $n=1,000$ participants, and the number of timepoints $m_{i}$ for each was multinomially distributed with equal probability of 1, 2, or 3 timepoints. The time between two measurements of a given participant was uniformly distributed between 1 and 6 years, which combined with the maximum number of 3 timepoints set the maximum possible follow-up interval to $(3-1) \times 6 = 12$ years. Baseline age was uniformly distributed between 4 and 90 years, and the date of initial measurement was uniformly distributed over 10 years, from 1st January 2000 to 1st January 2010. The simulations were repeated with identical dates of initial measurement. Random intercepts $b_{0i}$ and residuals $\epsilon_{ij}$ were sampled from normal distributions with mean zero and standard deviations equal to 50 \% and 20 \% of the sample standard deviation of the region's volume, respectively, similar to what was observed in the LCBC data. Datasets for each of the $12 \times 3 = 36$ combinations of regions and cohort effects were randomly sampled 1,000 times. 

Six models were fitted to each dataset, as summarized by Table \ref{tab:models}. The two formulations of model \eqref{eq:GAMM2} differ in that for (2b), the term $f(a_{i1},t_{ij})$ is a smooth bivariate function of $a_{i1}$ and $t_{ij}$ defined through the tensor product construction of \citet{Wood2012b} (see the left plot in Figure \ref{fig:mod2b} for an illustration), whereas (2a) uses the stricter formulation $f(a_{i1},t_{ij}) = f(a_{i1}) + \beta(a_{i1})t_{ij}$, where $f(a_{i1})$ is the main effect of age and $\beta(a_{i1})$ is a varying-coefficient term \citep{Hastie1993} representing the effect of time as a function of age. Model (2a) thus assumes that the effect of time is linear, with a slope that depends on baseline age. Models (3a) and (3b) differ in that (3b) allows the cohort effect to depend on age, with a term $\beta(a_{ij})c_{i}$ as shown in \eqref{eq:GAMM3}, while (3a) assumes that the cohort effect is age-independent, i.e., $\beta(a_{ij}) = \beta$ for all $a_{ij}$. See Supplementary Section S2.1 for precise mathematical definitions of the models.

\begin{table}[ht]
\centering
\begin{tabularx}{\textwidth}{lX}
\toprule
Identifier & Description \\
\midrule
(1a) & Model \eqref{eq:GAMM1} without random effects, using only the first timepoint.\\
(1b) & Model \eqref{eq:GAMM1} fitted to the complete data.\\
(2a) & Model \eqref{eq:GAMM2} with a varying-coefficient term for the interaction between baseline age and time.\\
(2b) & Model \eqref{eq:GAMM2} with a two-dimensional smooth function for jointly modeling the effect of baseline age and time.\\
(3a) & Model \eqref{eq:GAMM3} with linear age-independent cohort effects. \\
(3b) & Model \eqref{eq:GAMM3} with a varying-coefficient term allowing cohort-age interactions.\\
\bottomrule
\end{tabularx}
\caption{Models used in simulation experiments with GAMMs. The 'Identifier' column describes the name used to identify the model in the simulation results presented in Section \ref{sec:simulations} and in the Supplementary Material.}
\label{tab:models}
\end{table}

\section{Results}
\label{sec:results}

\subsection{Simulation experiments}
\label{sec:simulations}

\begin{figure}
\centering
\includegraphics[width=\columnwidth]{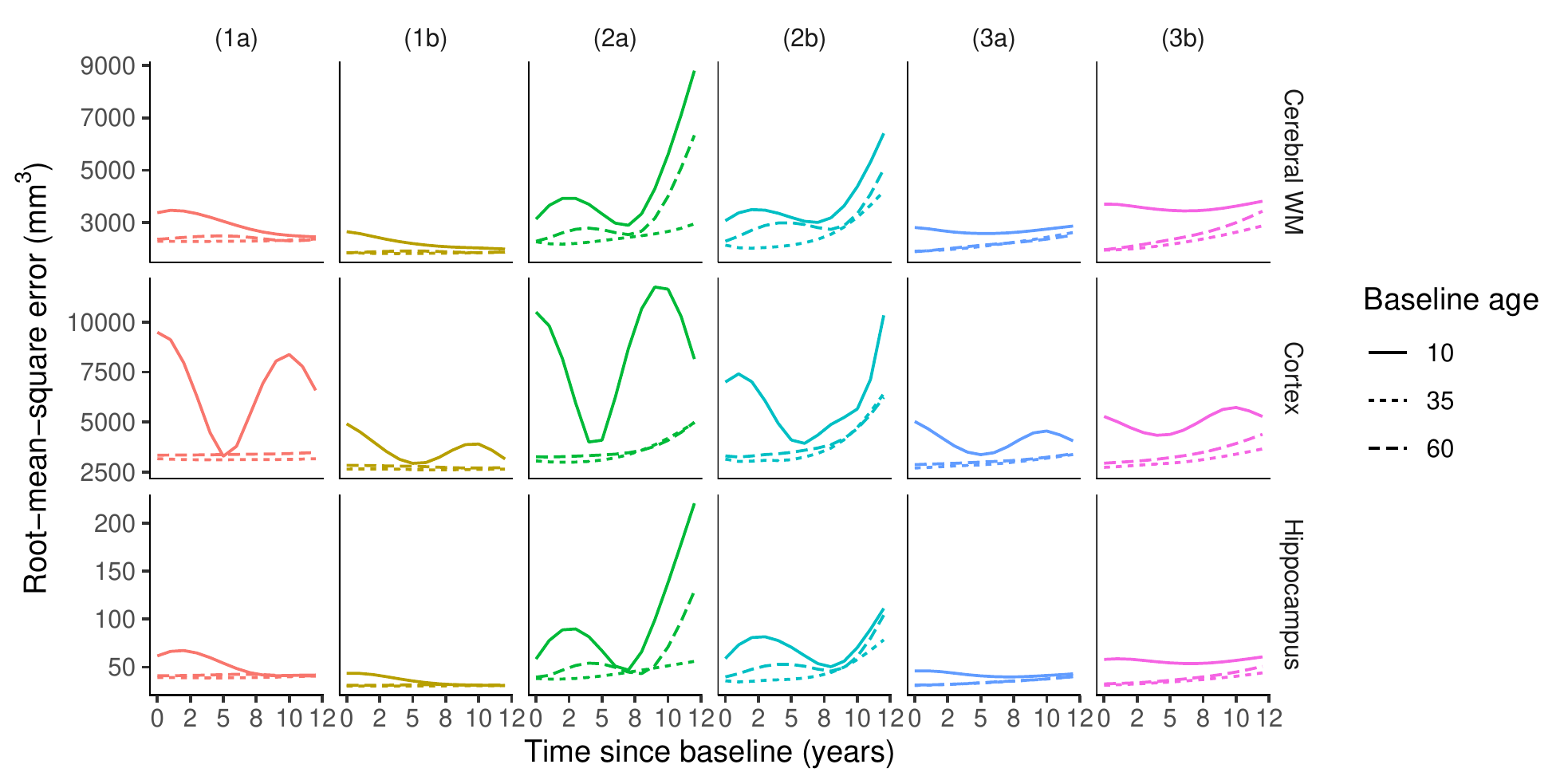}
\caption{\textbf{Longitudinal estimates with no cohort effects}. Simulation results in the case of no cohort effects, showing the RMSE of the predicted value after baseline ages 10, 35, and 60 years. For any given time $t$ along the $x$-axis, the curves represent the RMSE of the predicted longitudinal effect of $t$ years of increased age since baseline. Column headers specify the model fitted to the data, as defined in Table \ref{tab:models}.}
\label{fig:simres_long_rmse_summary_varying_date_no_cohort}
\end{figure}

\begin{figure}
\centering
\includegraphics[width=\columnwidth]{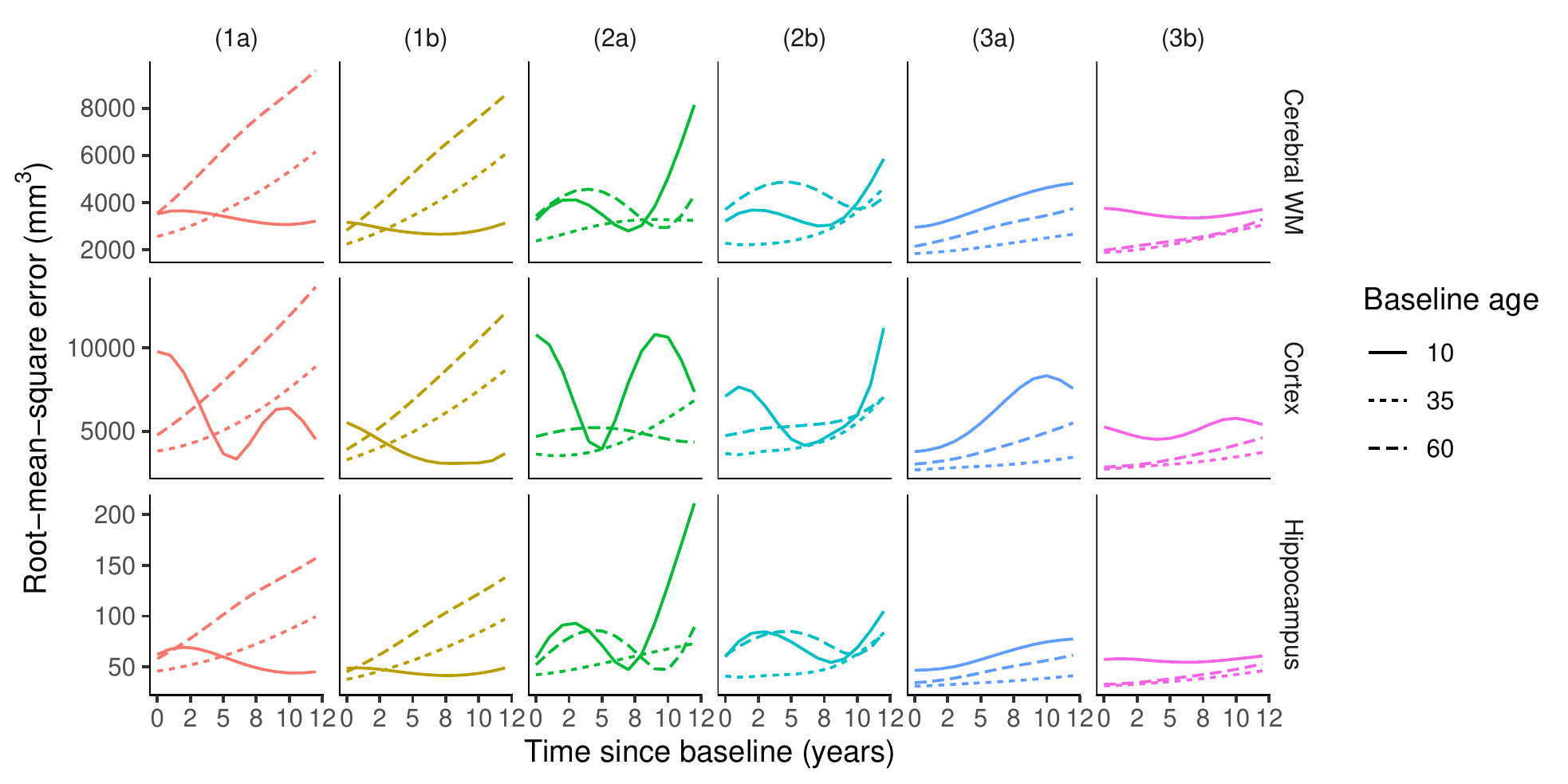}
\caption{\textbf{Longitudinal estimates with cohort-age interactions}. Simulation results in the case of age-cohort interactions, showing the RMSE of the predicted value after baseline ages 10, 35, and 60 years. For any given time $t$ along the $x$-axis, the curves represent the RMSE of the predicted longitudinal effect of $t$ years of increased age since baseline. Column headers specify the model fitted to the data, as defined in Table \ref{tab:models}.}
\label{fig:simres_long_rmse_summary_varying_date_cohort_interactions}
\end{figure}

Figure \ref{fig:simres_long_rmse_summary_varying_date_no_cohort} shows root-mean-square error (RMSE) of longitudinal estimates for each of the first 12 years after baseline ages 10, 35, and 60 years in the case of no cohort effects. Overall, model \eqref{eq:GAMM1} with longitudinal data had the most accurate fits, but the two variants of model \eqref{eq:GAMM3} were close. The two variants of model \eqref{eq:GAMM2}, on the other hand, had poorer fits than the other models, even for times very close after baseline, for which the data contained a large number of observations. Figure \ref{fig:simres_long_rmse_summary_varying_date_cohort_interactions} shows the results in the presence of age-cohort interactions. Now model \eqref{eq:GAMM1} with or without longitudinal data had higher RMSE than model \eqref{eq:GAMM2} for baseline ages 35 and 60 years, but lower RMSE for baseline age 10 years. Model \eqref{eq:GAMM3} had the lowest RMSE for all baseline ages in this case. Model (3b), which allows cohort-age interactions, had better overall performance than model (3a), which only contains the cohort effect as a single offset term. Results for age-independent cohort effects and for other regions were similar, and are shown in Supplementary Section S2.3.1.

Table \ref{tab:simres} shows the RMSE of the longitudinal estimates 12 years ahead averaged over each year and over baseline ages of 10, 35, and 60 years, along with its bias-variance decomposition $\text{RMSE}^{2} = \text{Bias}^{2} + \text{Variance}$ (e.g., \citet{Hastie2008}). The bias here quantifies the systematic error made by the model at any given baseline age and time, while the variance represents how much the model fit differs from one dataset to another\footnote{Table \ref{tab:simres} shows the square root of the average squared bias and the square root of the average variance, but for ease of presentation we will refer to these quantities as "bias" and "variance", respectively.}. In the absence of cohort effects, model (1b) utilizing longitudinal data had the lowest RMSE and variance across all regions. The two formulations of model \eqref{eq:GAMM3} had bias close to (1b), but higher variance. This can be attributed to the fact that the cohort terms in (3a) and (3b) are unnecessary in this case, and increase the variance because they increase the number of parameters to be estimated. Models (2a) and (2b) had the highest RMSE and bias for all regions in the absence of cohort effects. With age-independent cohort effects, model (3a) which includes the cohort effect as a single offset term had lower RMSE and bias than the other models across all regions. In this case, model (1b) had the lowest variance, at the cost of a much higher bias than models (3a) and (3b). The two formulations of model \eqref{eq:GAMM1} and the two formulations of model \eqref{eq:GAMM2} had considerably higher RMSE than either formulation of model \eqref{eq:GAMM3}, for all regions. Finally, with cohort-age interactions, model (3b) had lower bias and RMSE than the other models for all regions. Model (2b), which contains the sufficient terms to capture such an interaction effects, had higher RMSE than both versions of model \eqref{eq:GAMM3} across all regions. In this case the two formulations of model \eqref{eq:GAMM1} also performed poorly.

\begin{table}
\centering
\resizebox{\linewidth}{!}{
\input{tables/simres_varying_date.txt}
}
\caption{\label{tab:simres}RMSE, bias, and variance of longitudinal estimates averaged over the next 12 years following baseline ages 10, 35, and 60 years. Mean-square error, squared bias, and variance of the prediction were averaged over all Monte Carlo samples for each baseline age and time, and then averaged over baseline ages and times. The square roots of these averages are shown for each region, model, and cohort effect. That is, each cell in column 3 (no cohort effect, RMSE) represents the average of a subplot in Figure \ref{fig:simres_long_rmse_summary_varying_date_no_cohort}, and similarly each cell in column 9 (cohort-age interaction, RMSE) represents the average of a subplot in Figure \ref{fig:simres_long_rmse_summary_varying_date_cohort_interactions}.}
\end{table}

\begin{figure}
\centering
\includegraphics{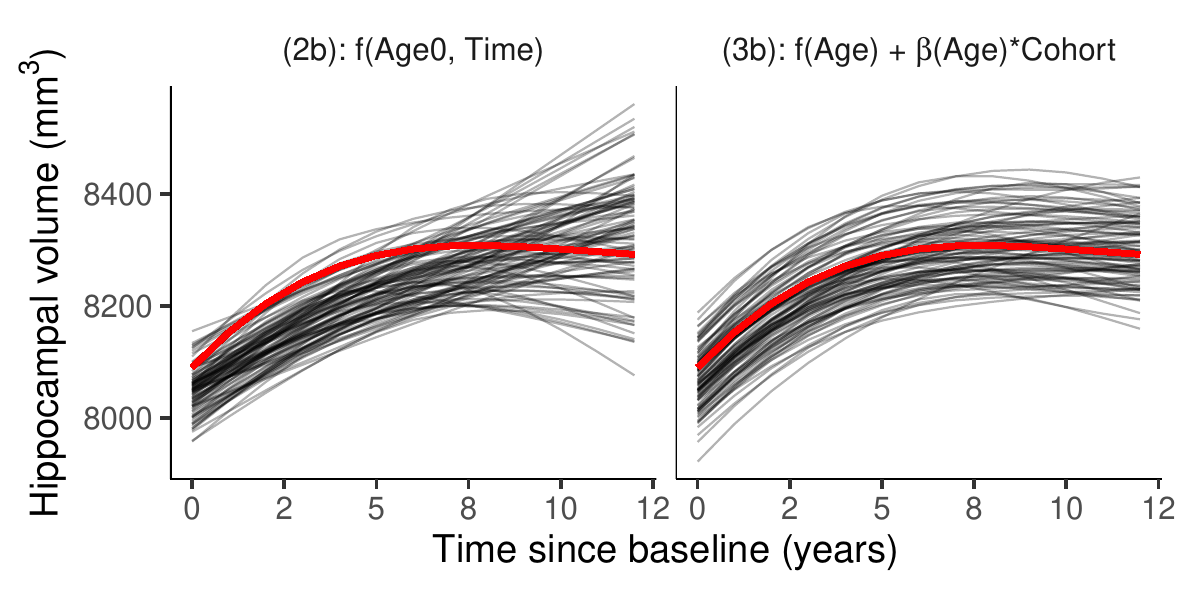}
\caption{\textbf{Sample fits}. A random sample of 100 fits in the case of cohort-age interaction for hippocampal volume. Thin lines show estimated longitudinal effects from baseline age 10, and the thick red lines show the true values.}
\label{fig:hippo_long_samples_varying_date}
\end{figure}

Figure \ref{fig:hippo_long_samples_varying_date} shows estimated trajectories of hippocampal volume from a baseline age 10 years, from a random subset of 100 out of the total 1,000 models fitted to the simulated datasets. While the estimates of model \eqref{eq:GAMM3} follow the true effect over the full 12-year period, a large proportion of the estimates of model \eqref{eq:GAMM2} are close to straight lines. Due to the simulated dropout after the first or second timepoint, combined with the unstructured time intervals between measurements, the average follow-up time in the data is only 3.5 years, and thus much lower than the maximum follow-up of 12 years. A consequence of this data structure, which resembles the LCBC data shown in Figure \ref{fig:date_dist}, is that for the two formulations of model \eqref{eq:GAMM2}, there is not enough data to estimate the effect of time beyond the first few years after baseline, even though the maximum follow-up interval is 12 years. With a limited amount of data for timepoints further than 3-4 years from baseline, the second derivative penalization used by GAMMs pulls the estimates towards straight lines, which have zero second derivatives and hence are not penalized, an effect which is clearly seen in Figure \ref{fig:hippo_long_samples_varying_date}.

Simulation results with identical baseline dates shown in Supplementary Section S2.4 are practically identical to those described in this section, suggesting that the issue of varying baseline dates is not critical for GAMMs when the variation maximum variation is ten years. Instead, as Figure \ref{fig:hippo_long_samples_varying_date} shows, the main challenge with estimating longitudinal effects using the GAMM \eqref{eq:GAMM2} is caused by the fact that time $t_{ij}$ for most participants spans a short period compared to the full follow-up interval, making estimation of nonlinear effects increasingly challenging as time since baseline increases. For estimation of cross-sectional effects, the differences between models were smaller. However, the two versions of model \eqref{eq:GAMM2} still showed the poorest performance, while model (1b) and the two versions of model \eqref{eq:GAMM3} showed the best performance. Detailed results for estimation of cross-sectional effects are shown in Supplementary Section S2.3.2.

\subsection{Example Applications}
\label{sec:UseCases}

In this section we show how GAMMs can be applied to the study of lifespan brain development, with example \verb!R! code using the packages \verb!mgcv! \citep{Wood2017} and \verb!gamm4! \citep{Wood2017gamm4}. Some parts of the code are omitted for ease of presentation, and can be found in the Supplementary Section S3.

\subsubsection{Modeling Lifespan Volume Trajectories}
\label{sec:LifespanHippovol}

We first consider the hippocampal volumes shown in Figure \ref{fig:spaghetti_plots} (right). The data are organized in long format in the dataframe \verb!dat!, with each row representing one timepoint of a single participant, containing the following variables:

\begin{itemize}[noitemsep]
\item \verb!ID!: Unique participant ID.
\item \verb!Age!: Age in years at MRI session.
\item \verb!Hippocampus!: Estimated hippocampal volume in mm$^3$.
\item \verb!ICV_z!: Estimated total intracranial volume, standardized to have zero mean and unit variance.
\item \verb!Sex!: Participant sex, coded as "Female" and "Male".
\item \verb!Scanner!: Factor variable indicating which scanner was used for MRI.
\item \verb!Age_bl!: Age in years at initial MRI session.
\item \verb!Time!: Time in years since initial MRI session.
\item \verb!Birth_Date_z!: Decimal number of years between birth date and 1st January 1970.
\end{itemize}

The transformed variables with suffix \verb!_z! were created because the algorithms used in the models to be fitted are most stable when the explanatory variables are of similar magnitude.

\paragraph{Models not separating longitudinal and cross-sectional effects}

We start by fitting a GAM with only the first timepoint of each participant, using the \verb!gam()! function from \verb!mgcv!. By default, \verb!gam()! uses generalized cross-validation \citep{Golub1979} for smoothing, but for comparison with mixed models we specify that restricted maximum likelihood (REML) should be used, with the argument \verb!method = "REML"!. The smooth function corresponding to the term $f(a_{ij})$ in model \eqref{eq:GAMM1} is specified with \verb!s(Age, k = 20, bs = "cr")!, where we use \verb!k = 20! cubic regression (\verb!bs = "cr"!) splines. The default is thin-plate splines (\verb!bs = "tp"!), but in our experience cubic regression splines typically require half the computing time, without yielding poorer fit. ICV, sex, and scanner are added as additional covariates. Throughout this section, the names of the fitted models correspond to the model identifiers in Table \ref{tab:models}.

\begin{knitrout}
\definecolor{shadecolor}{rgb}{0.969, 0.969, 0.969}\color{fgcolor}\begin{kframe}
\begin{alltt}
\hlkwd{library}\hlstd{(mgcv)}
\hlcom{# Keep only first timepoint (Time = 0)}
\hlstd{baseline_data} \hlkwb{<-} \hlkwd{subset}\hlstd{(dat, Time} \hlopt{==} \hlnum{0}\hlstd{)}
\hlcom{# Fit GAM to data}
\hlstd{mod1a} \hlkwb{<-} \hlkwd{gam}\hlstd{(Hippocampus} \hlopt{~} \hlkwd{s}\hlstd{(Age,} \hlkwc{k} \hlstd{=} \hlnum{20}\hlstd{,} \hlkwc{bs} \hlstd{=} \hlstr{"cr"}\hlstd{)} \hlopt{+} \hlstd{ICV_z} \hlopt{+} \hlstd{Sex} \hlopt{+} \hlstd{Scanner,}
             \hlkwc{data} \hlstd{= baseline_data,} \hlkwc{method} \hlstd{=} \hlstr{"REML"}\hlstd{)}
\end{alltt}
\end{kframe}
\end{knitrout}

\begin{figure}
\centering
\includegraphics{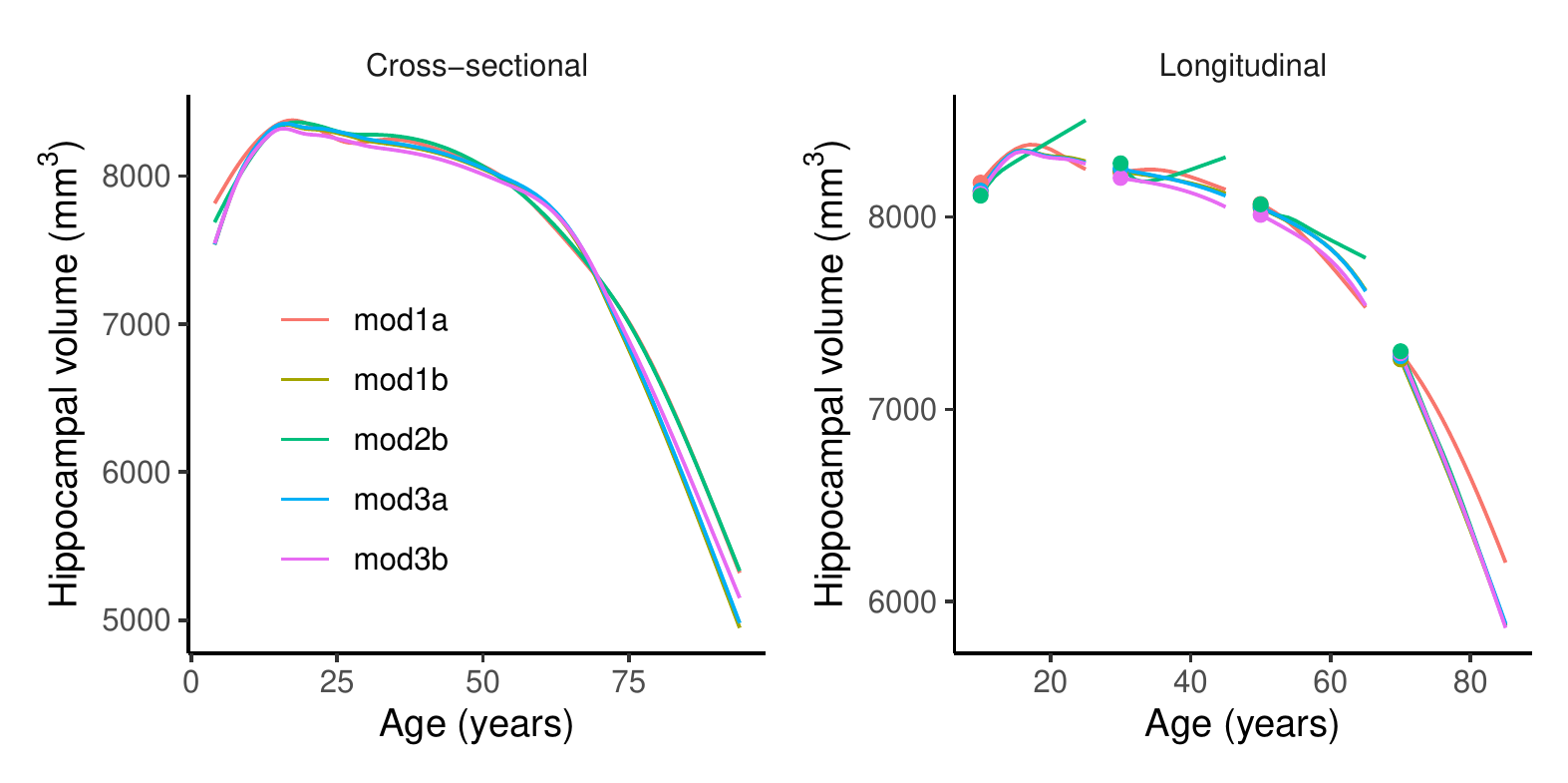}
\caption{\textbf{Model estimates}. Estimated cross-sectional and longitudinal effects from each of the five models considered in Section \ref{sec:LifespanHippovol}. The cross-sectional estimates are computed for 1st January 2010. The longitudinal estimates are computed 15 years ahead from baseline ages of 10, 30, 50, and 70 years.}
\label{fig:cross_long_effects_hippocampus_example}
\end{figure}

The estimated smooth function can be immediately visualized with the \verb!plot()! function. The output is not shown, but see the curve labeled \verb!mod1a! in Figure \ref{fig:cross_long_effects_hippocampus_example} (left).

\begin{knitrout}
\definecolor{shadecolor}{rgb}{0.969, 0.969, 0.969}\color{fgcolor}\begin{kframe}
\begin{alltt}
\hlkwd{plot}\hlstd{(mod1a)}
\end{alltt}
\end{kframe}
\end{knitrout}

We can check that the number of splines is sufficiently high with \verb!k.check()!, which implements a permutation algorithm described in \citet[Ch. 5.9]{Wood2017}. A significant $p$-value and estimated degrees of freedom (edf) close to the maximum degrees of freedom k', indicates that more splines are required. As shown in the output below, k seems sufficiently high. The maximum number of degrees freedom k' = 19 is one less than the number of cubic regression splines, because one degree of freedom is used to center the smooth function such that it has zero mean over the range of age values in the data \citep[Ch. 5.4.1]{Wood2017}.

\begin{knitrout}
\definecolor{shadecolor}{rgb}{0.969, 0.969, 0.969}\color{fgcolor}\begin{kframe}
\begin{alltt}
\hlkwd{k.check}\hlstd{(mod1a)}
\end{alltt}
\begin{verbatim}
##        k'      edf   k-index p-value
## s(Age) 19 7.862725 0.9946078    0.38
\end{verbatim}
\end{kframe}
\end{knitrout}

Next, we fit model \eqref{eq:GAMM1} using the complete data. This can be achieved both with the \verb!gam()! and \verb!gamm()! functions from \verb!mgcv!, and the \verb!gamm4()! function from \verb!gamm4!. For the type of longitudinal data considered here, with a low number of repeated measurements of a large number of participants, \verb!gam()! is very slow compared to \verb!gamm()! and \verb!gamm4()!. We opt for the latter, as it is typically faster and more numerically stable. The main difference from the model with only cross-sectional data, is that we now specify a random intercept with the argument \verb!random = ~(1|ID)!.

\begin{knitrout}
\definecolor{shadecolor}{rgb}{0.969, 0.969, 0.969}\color{fgcolor}\begin{kframe}
\begin{alltt}
\hlkwd{library}\hlstd{(gamm4)}
\hlstd{mod1b} \hlkwb{<-} \hlkwd{gamm4}\hlstd{(Hippocampus} \hlopt{~} \hlkwd{s}\hlstd{(Age,} \hlkwc{k} \hlstd{=} \hlnum{20}\hlstd{,} \hlkwc{bs} \hlstd{=} \hlstr{"cr"}\hlstd{)} \hlopt{+} \hlstd{ICV_z} \hlopt{+} \hlstd{Sex} \hlopt{+} \hlstd{Scanner,}
               \hlkwc{data} \hlstd{= dat,} \hlkwc{random} \hlstd{=} \hlopt{~}\hlstd{(}\hlnum{1}\hlopt{|}\hlstd{ID))}
\end{alltt}
\end{kframe}
\end{knitrout}

On a MacBook Pro, fitting this GAMM took 1.7 seconds, while fitting the GAM above took less than 0.08 seconds. The \verb!gamm4()! function returns a list with two elements, named \verb!mer! and \verb!gam!. The element \verb!mer! contains information from the \verb!lmer()! function in the \verb!lme4! package \citep{Bates2015} used in the numerical computations, and is useful for studying the random effect distributions. The element \verb!gam! contains information about the smooth functions, and is useful for studying smooth terms and parametric fixed effects.

\begin{knitrout}
\definecolor{shadecolor}{rgb}{0.969, 0.969, 0.969}\color{fgcolor}\begin{kframe}
\begin{alltt}
\hlcom{# Print information about model object:}
\hlkwd{str}\hlstd{(mod1b,} \hlkwc{max.level} \hlstd{=} \hlnum{1}\hlstd{)}
\end{alltt}
\begin{verbatim}
## List of 2
##  $ mer:Formal class 'lmerMod' [package "lme4"] with 13 slots
##  $ gam:List of 32
##   ..- attr(*, "class")= chr "gam"
\end{verbatim}
\end{kframe}
\end{knitrout}

The \verb!lme4! package is automatically loaded when \verb!gamm4! is loaded, and its accessor functions can be used to study the random effect distributions. The summary below shows that the between-participant variation, $\hat{\sigma}_{b} = 601$ mm$^{3}$, is much larger than the within-participant variation, $\hat{\sigma} = 133$ mm$^{3}$. Note that for \verb!mod1a! which does not have random intercepts, this information is not available. The second line in the output ($\hat{\sigma}_{\lambda} = 21$ mm$^{3}$) is related to the formulation of smooth functions as random effects, and the estimated smoothing parameter is given by $\hat{\lambda} = \hat{\sigma}^2/\hat{\sigma}_{\lambda}^{2} = 38.7$.

\begin{knitrout}
\definecolor{shadecolor}{rgb}{0.969, 0.969, 0.969}\color{fgcolor}\begin{kframe}
\begin{alltt}
\hlkwd{VarCorr}\hlstd{(mod1b}\hlopt{$}\hlstd{mer)}
\end{alltt}
\begin{verbatim}
##  Groups   Name        Std.Dev.
##  ID       (Intercept) 601.378 
##  Xr       s(Age)       21.457 
##  Residual             133.410
\end{verbatim}
\end{kframe}
\end{knitrout}

\paragraph{Modeling cohort effects}
\label{sec:cohort_effects_varcoef}

Next, we take cohort effects into account by fitting two versions of model \eqref{eq:GAMM3}, \verb!mod3a! which contains a linear cohort effect term, and \verb!mod3b! which contains a varying-coefficient term $\beta(a_{ij})$ consisting of five cubic regression splines, allowing the cohort effect to depend on age.

\begin{knitrout}
\definecolor{shadecolor}{rgb}{0.969, 0.969, 0.969}\color{fgcolor}\begin{kframe}
\begin{alltt}
\hlstd{mod3a} \hlkwb{<-} \hlkwd{gamm4}\hlstd{(Hippocampus} \hlopt{~} \hlkwd{s}\hlstd{(Age,} \hlkwc{k} \hlstd{=} \hlnum{20}\hlstd{,} \hlkwc{bs} \hlstd{=} \hlstr{"cr"}\hlstd{)} \hlopt{+}
                 \hlstd{Birth_Date_z} \hlopt{+} \hlstd{ICV_z} \hlopt{+} \hlstd{Sex} \hlopt{+} \hlstd{Scanner,} \hlkwc{data} \hlstd{= dat,} \hlkwc{random} \hlstd{=} \hlopt{~}\hlstd{(}\hlnum{1}\hlopt{|}\hlstd{ID))}

\hlstd{mod3b} \hlkwb{<-} \hlkwd{gamm4}\hlstd{(Hippocampus} \hlopt{~} \hlkwd{s}\hlstd{(Age,} \hlkwc{k} \hlstd{=} \hlnum{20}\hlstd{,} \hlkwc{bs} \hlstd{=} \hlstr{"cr"}\hlstd{)} \hlopt{+}
                \hlkwd{s}\hlstd{(Age,} \hlkwc{by} \hlstd{= Birth_Date_z,} \hlkwc{bs} \hlstd{=} \hlstr{"cr"}\hlstd{,} \hlkwc{k} \hlstd{=} \hlnum{5}\hlstd{)} \hlopt{+} \hlstd{ICV_z} \hlopt{+} \hlstd{Sex} \hlopt{+} \hlstd{Scanner,}
               \hlkwc{data} \hlstd{= dat,} \hlkwc{random} \hlstd{=} \hlopt{~}\hlstd{(}\hlnum{1}\hlopt{|}\hlstd{ID))}
\end{alltt}
\end{kframe}
\end{knitrout}

We extract the cohort effect estimated by \verb!mod3a! from the matrix \verb!p.table! in the object returned by \verb!mgcv!'s \verb!summary()! function. A 95 \% confidence interval (CI) is computed by adding the standard error multiplied by the 2.5 \% and 97.5 \% quantiles of the standard normal distribution to the estimate. As the output shows, the estimated cohort effect from this model is a negative offset of 1.25 mm$^3$ per birth year, with 95 \% CI $[-4.29, 1.78]$ mm$^3$.

\begin{knitrout}
\definecolor{shadecolor}{rgb}{0.969, 0.969, 0.969}\color{fgcolor}\begin{kframe}
\begin{alltt}
\hlcom{# Extract birth date coefficient from table of parametric estimates}
\hlstd{coef_info} \hlkwb{<-} \hlkwd{summary}\hlstd{(mod3a}\hlopt{$}\hlstd{gam)}\hlopt{$}\hlstd{p.table[}\hlstr{"Birth_Date_z"}\hlstd{, ]}

\hlcom{# Compute 95 % confidence intervals and add them to coef_info}
\hlstd{confints} \hlkwb{<-} \hlkwd{sapply}\hlstd{(}\hlkwd{c}\hlstd{(}\hlkwc{conf.low} \hlstd{=} \hlnum{.025}\hlstd{,} \hlkwc{conf.high} \hlstd{=} \hlnum{.975}\hlstd{),} \hlkwa{function}\hlstd{(}\hlkwc{p}\hlstd{)}
  \hlstd{coef_info[[}\hlstr{"Estimate"}\hlstd{]]} \hlopt{+} \hlkwd{qnorm}\hlstd{(p)} \hlopt{*} \hlstd{coef_info[[}\hlstr{"Std. Error"}\hlstd{]])}

\hlstd{(coef_info} \hlkwb{<-} \hlkwd{c}\hlstd{(coef_info, confints))}
\end{alltt}
\begin{verbatim}
##   Estimate Std. Error    t value   Pr(>|t|)   conf.low  conf.high 
##    -1.2544     1.5468    -0.8109     0.4175    -4.2862     1.7774
\end{verbatim}
\end{kframe}
\end{knitrout}

For example, since the variable \verb!Birth_Date_z! represents the number of years between the participant's birth date and 1st January 1970, we can estimate the offset effect of being born in 1970 compared to being born in 1920 by multiplying \verb!coef_info! by 50 years\cprotect\footnote{Since the cohort effect estimated by \verb!mod3a! does not interact with age, the result applies to any set of birth dates separated by 50 years.}. This is illustrated in the line below, which shows that the estimate is $-62.7$ mm$^3$ with 95 \% CI $[-214, 89.0]$ mm$^3$. The upper and lower limits of the CI are small, but not negligible compared to the sample average of 8,065 mm$^3$.

\begin{knitrout}
\definecolor{shadecolor}{rgb}{0.969, 0.969, 0.969}\color{fgcolor}\begin{kframe}
\begin{alltt}
\hlstd{coef_info[}\hlkwd{c}\hlstd{(}\hlstr{"Estimate"}\hlstd{,} \hlstr{"conf.low"}\hlstd{,} \hlstr{"conf.high"}\hlstd{)]} \hlopt{*} \hlnum{50}
\end{alltt}
\begin{verbatim}
##  Estimate  conf.low conf.high 
##    -62.72   -214.31     88.87
\end{verbatim}
\end{kframe}
\end{knitrout}

Next, the varying-coefficient term in \verb!mod3b! is extracted from the matrix \verb!s.table! in the object returned by \verb!mgcv!'s \verb!summary()! function. Its estimated degrees of freedom is 2, implying that the cohort effect is estimated as a straight line defined by an intercept and a slope. Its $p$-value of $0.0506$ also suggests that there is some evidence of an age-dependent cohort effect.

\begin{knitrout}
\definecolor{shadecolor}{rgb}{0.969, 0.969, 0.969}\color{fgcolor}\begin{kframe}
\begin{alltt}
\hlkwd{summary}\hlstd{(mod3b}\hlopt{$}\hlstd{gam)}\hlopt{$}\hlstd{s.table[}\hlstr{"s(Age):Birth_Date_z"}\hlstd{, ]}
\end{alltt}
\begin{verbatim}
##     edf  Ref.df       F p-value 
## 2.00000 2.00000 2.98508 0.05064
\end{verbatim}
\end{kframe}
\end{knitrout}

The estimated cohort effect can be plotted with the code shown below, using the \verb!plot()! function for \verb!gam! objects. This term is numbered 2 because it was entered as the second smooth term in the formula specifying the model. The argument \verb!scale = 0! ensures that the y-axis limits are adjusted to the term to be plotted, rather than also covering the full range of the term representing the main effect of age.

\begin{knitrout}
\definecolor{shadecolor}{rgb}{0.969, 0.969, 0.969}\color{fgcolor}\begin{kframe}
\begin{alltt}
\hlkwd{plot}\hlstd{(mod3b}\hlopt{$}\hlstd{gam,} \hlkwc{select} \hlstd{=} \hlnum{2}\hlstd{,} \hlkwc{scale} \hlstd{=} \hlnum{0}\hlstd{)}
\end{alltt}
\end{kframe}
\end{knitrout}

\begin{figure}
\centering
\includegraphics{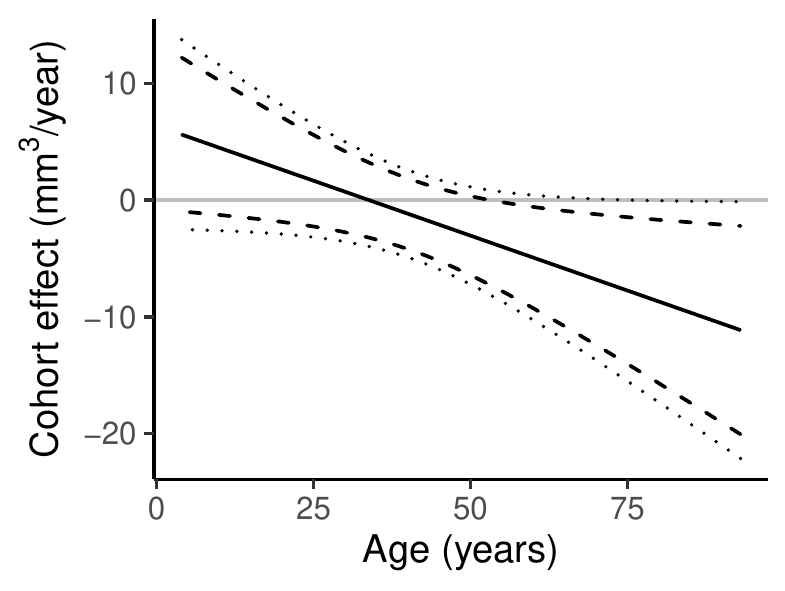}
\caption{\textbf{Estimated cohort effect}. Estimated cohort effect on hippocampal volume as a function of age. Dashed lines show 95 \% across-the-function CIs, which have the property that the true function is expected to lie within the CI over 95 \% of the $x$-axis. Dotted lines show 95 \% simultaneous CIs, which have the property that the true function is expected to be completely confined within the CI 95 \% of the time under repeated sampling from the population.}
\label{fig:varying_coefficient_hippocampus}
\end{figure}

A slightly modified version of the resulting plot is shown by the solid and dashed lines in Figure \ref{fig:varying_coefficient_hippocampus}. The fact that the estimated cohort effect averaged over all ages is slightly negative is in agreement with \verb!mod3a!, which estimated a negative but non-significant cohort effect. The CIs shown by the \verb!plot()! function have the property that under repeated sampling from the population, the true function will on average be confined within the upper and lower limits over 95 \% of the $x$-axis \citep{Marra2012,Nychka1988}. These across-the-function CIs will contain the true function less than 95 \% of the time under repeated sampling from the population, which explains why the upper limit in Figure \ref{fig:varying_coefficient_hippocampus} is well below zero despite the $p$-value being larger than 0.05. Simultaneous CIs, on the other hand, would fully contain the complete function 95 \% of the time under repeated sampling, and can be constructed using a simulation-based approach \citep{Ruppert2003,Simpson2016} shown in the Supplementary Section S3.1. These simultaneous CIs are shown as the dotted lines in Figure \ref{fig:varying_coefficient_hippocampus}, and are wider than the across-the-function CIs. The fact that the upper limit is very close to zero for high ages and the lower limit never is above zero, is in agreement with the $p$-value being approximately 0.05.

The age-dependent cohort effects estimated by \verb!mod3b! imply that a participant born in 1920 is expected to have a 131 mm$^3$ lower hippocampal volume at age 20 than a participant born in 1970 at age 20 (CI: $[-407, 145]$ mm$^3$). Conversely, a participant born in 1920 is expected to have a 340 mm$^3$ higher hippocampal volume than a participant born in 1970, at age 70 (CI: $[-5.8, 685]$ mm$^3$). R code for computing these estimates is shown in Supplementary Section S3.1.1. As for \verb!mod3a!, these results suggest that a cohort effect cannot be ruled out, despite the term not being significant, since cohort effects of relatively large magnitude are contained within the 95 \% CIs.

\paragraph{Modeling baseline age and time since baseline}

Finally, we fit model \eqref{eq:GAMM2}, using the \verb!t2()! function to create a two-dimensional smooth term \citep{Wood2012b}. The argument \verb!k = c(20, 5)! specifies that 20 cubic regression splines are used for the effect of baseline age, while only 5 are used for the effect of time, as this term does not span more than 11 years. The construction of the two-dimensional function involves forming products of all combinations of splines for baseline age and time, implying that the total number of degrees of freedom used by the term equals $20 \times 5 - 1 = 99$, where the one degree of freedom subtracted has been used for imposing a sum-to-zero constraint. Fitting \verb!mod2b! below took $\approx$ 90 seconds on a MacBook Pro.

\begin{knitrout}
\definecolor{shadecolor}{rgb}{0.969, 0.969, 0.969}\color{fgcolor}\begin{kframe}
\begin{alltt}
\hlstd{mod2b} \hlkwb{<-} \hlkwd{gamm4}\hlstd{(Hippocampus} \hlopt{~} \hlkwd{t2}\hlstd{(Age_bl, Time,} \hlkwc{k} \hlstd{=} \hlkwd{c}\hlstd{(}\hlnum{20}\hlstd{,} \hlnum{5}\hlstd{),} \hlkwc{bs} \hlstd{=} \hlstr{"cr"}\hlstd{)} \hlopt{+}
                \hlstd{ICV_z} \hlopt{+} \hlstd{Sex} \hlopt{+} \hlstd{Scanner,} \hlkwc{data} \hlstd{= dat,} \hlkwc{random} \hlstd{=} \hlopt{~}\hlstd{(}\hlnum{1}\hlopt{|}\hlstd{ID))}
\end{alltt}
\end{kframe}
\end{knitrout}

Model \eqref{eq:GAMM2} in Section \ref{sec:GAMMlong} could alternatively be formulated with three smooth terms: the main effects of age and time, and their interaction. This allows significance testing of each term separately.  Functionality for fitting such a model is provided by \verb!mgcv!'s \verb!ti()! function, representing a two-dimensional tensor interaction term in which the main effects have been removed \citep{Wood2006}. As it is not available in \verb!gamm4!, the \verb!gamm()! function needs to be used. The syntax is very similar, except that the random intercept is specified with \verb!random = list(ID =~ 1)! and the use of REML rather than the default marginal maximum likelihood is specified with \verb!method = "REML"! for comparability with the models fitted with \verb!gamm4()!.

\begin{knitrout}
\definecolor{shadecolor}{rgb}{0.969, 0.969, 0.969}\color{fgcolor}\begin{kframe}
\begin{alltt}
\hlcom{# Alternative formulation with tensor interaction terms}
\hlstd{mod2b_ti} \hlkwb{<-} \hlkwd{gamm}\hlstd{(Hippocampus} \hlopt{~} \hlkwd{s}\hlstd{(Age_bl,} \hlkwc{k} \hlstd{=} \hlnum{20}\hlstd{,} \hlkwc{bs} \hlstd{=} \hlstr{"cr"}\hlstd{)} \hlopt{+}
                   \hlkwd{s}\hlstd{(Time,} \hlkwc{k} \hlstd{=} \hlnum{5}\hlstd{,} \hlkwc{bs} \hlstd{=} \hlstr{"cr"}\hlstd{)} \hlopt{+}
                   \hlkwd{ti}\hlstd{(Age_bl, Time,} \hlkwc{k} \hlstd{=} \hlkwd{c}\hlstd{(}\hlnum{20}\hlstd{,} \hlnum{5}\hlstd{),} \hlkwc{bs} \hlstd{=} \hlstr{"cr"}\hlstd{)} \hlopt{+}
                   \hlstd{ICV_z} \hlopt{+} \hlstd{Sex} \hlopt{+} \hlstd{Scanner,}
                 \hlkwc{data} \hlstd{= dat,} \hlkwc{random} \hlstd{=} \hlkwd{list}\hlstd{(}\hlkwc{ID} \hlstd{=}\hlopt{~} \hlnum{1}\hlstd{),} \hlkwc{method} \hlstd{=} \hlstr{"REML"}\hlstd{)}
\end{alltt}
\end{kframe}
\end{knitrout}

Information about the smooth terms can again be obtained from \verb!s.table! returned by \verb!summary()!, and shows that all terms are significant. Interestingly, the main effect of time is estimated to be linear, as can be seen by its single degree of freedom, while the two-dimensional interaction term is highly nonlinear.

\begin{knitrout}
\definecolor{shadecolor}{rgb}{0.969, 0.969, 0.969}\color{fgcolor}\begin{kframe}
\begin{alltt}
\hlkwd{summary}\hlstd{(mod2b_ti}\hlopt{$}\hlstd{gam)}\hlopt{$}\hlstd{s.table}
\end{alltt}
\begin{verbatim}
##                    edf Ref.df      F p-value
## s(Age_bl)        7.648  7.648 138.26       0
## s(Time)          1.000  1.000  45.44       0
## ti(Age_bl,Time) 23.074 23.074  67.46       0
\end{verbatim}
\end{kframe}
\end{knitrout}

Two-dimensional smooth terms can also be visualized with the \verb!plot()! function, for which a perspective plot is produces by setting \verb!scheme = 1!. 

\begin{knitrout}
\definecolor{shadecolor}{rgb}{0.969, 0.969, 0.969}\color{fgcolor}\begin{kframe}
\begin{alltt}
\hlkwd{plot}\hlstd{(mod2b}\hlopt{$}\hlstd{gam,} \hlkwc{scheme} \hlstd{=} \hlnum{1}\hlstd{)} \hlcom{# Plot full tensor product}
\hlkwd{plot}\hlstd{(mod2b_ti}\hlopt{$}\hlstd{gam,} \hlkwc{select} \hlstd{=} \hlnum{3}\hlstd{,} \hlkwc{scheme} \hlstd{=} \hlnum{1}\hlstd{)} \hlcom{# Plot tensor interaction, term #3}
\end{alltt}
\end{kframe}
\end{knitrout}

\begin{figure}
\centering
\includegraphics[width=.49\columnwidth]{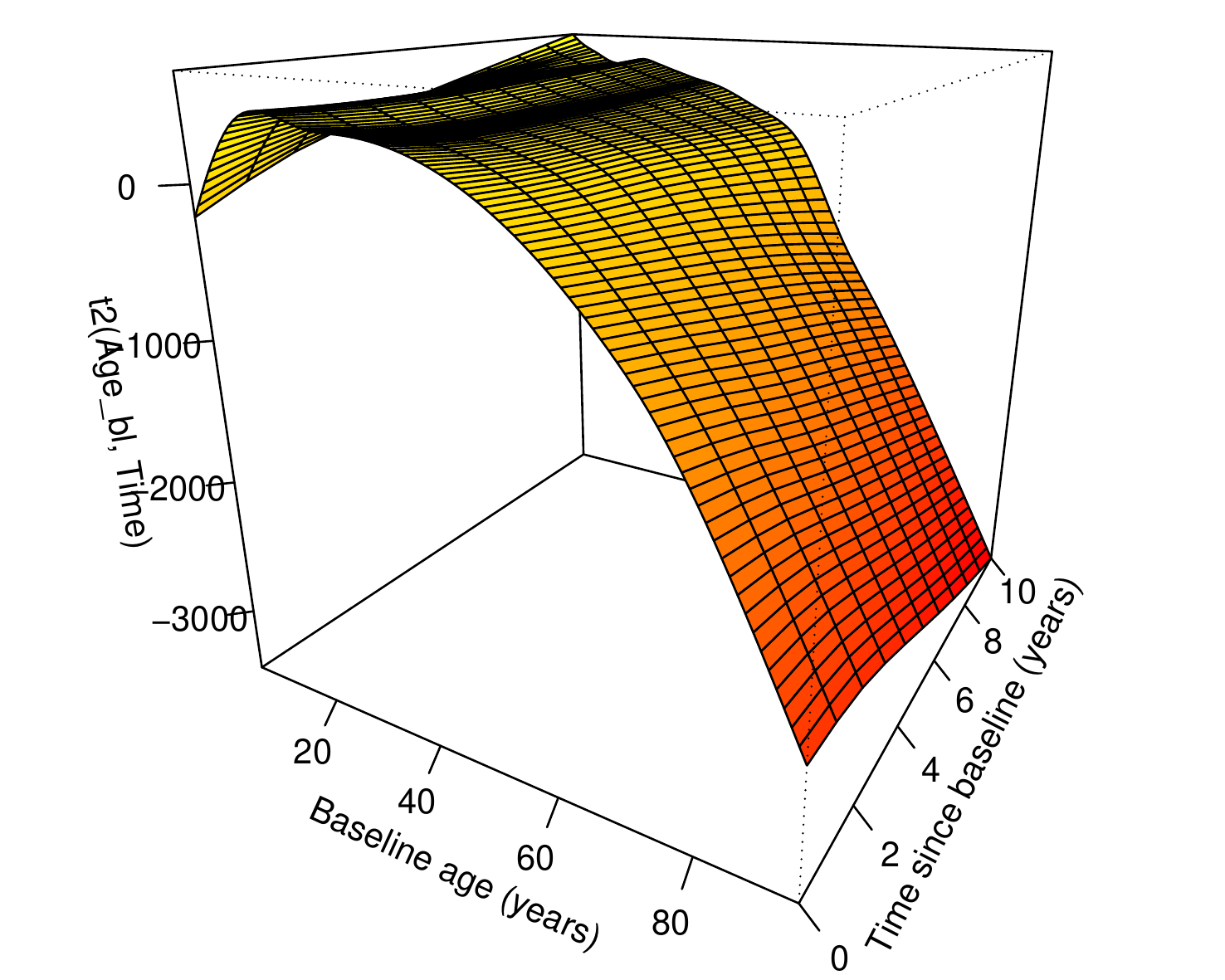}
\includegraphics[width=.49\columnwidth]{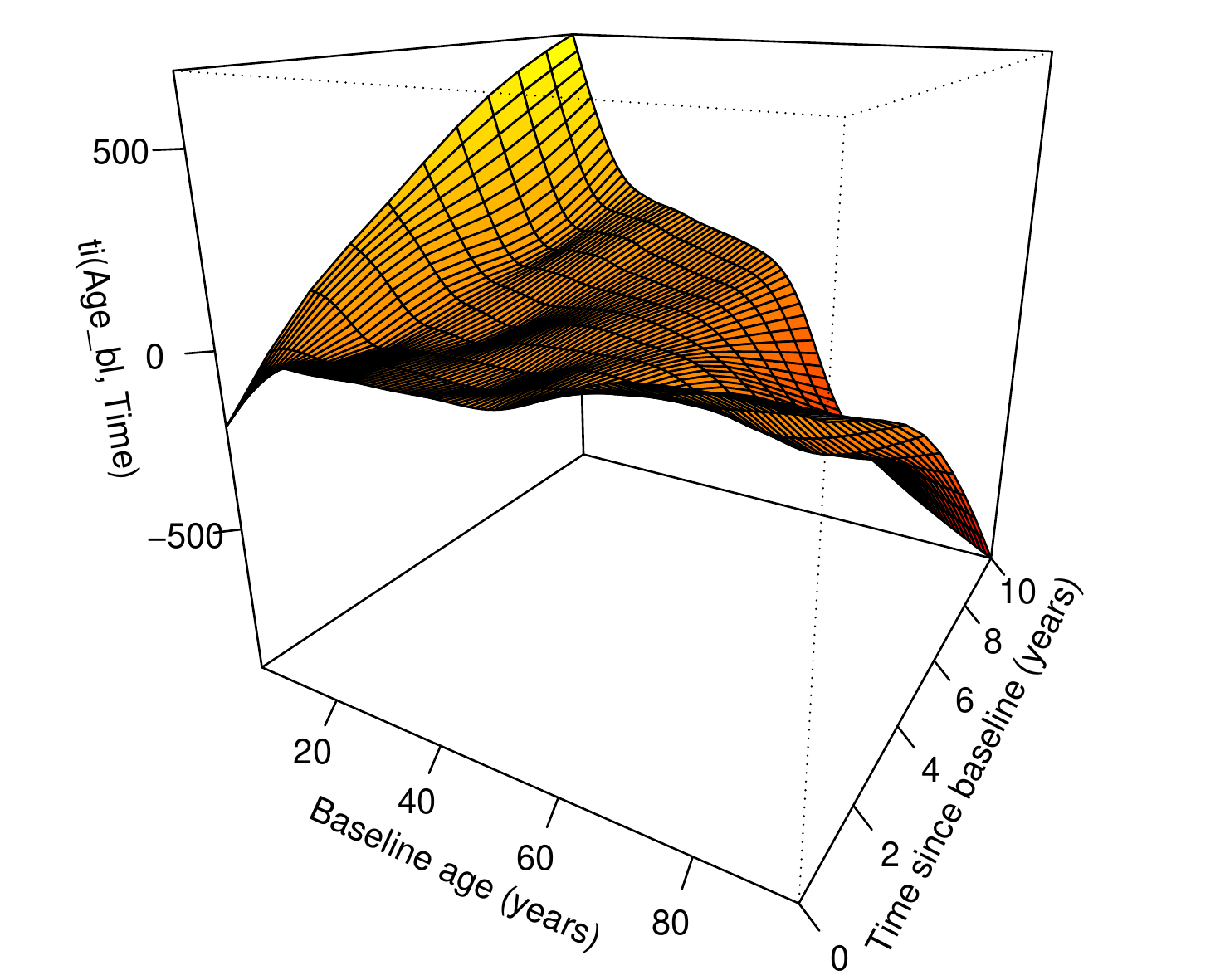}
\cprotect\caption{\textbf{Two-dimensional smooth functions}. Left: tensor product term \verb!t2(Age_bl, Time)! in \verb!mod2b!, representing the total effect of baseline age and time. Right: tensor interaction term \verb!ti(Age_bl, Time)! in \verb!mod2b_ti!, representing the interaction effect between baseline age and time.}
\label{fig:mod2b}
\end{figure}

The resulting plots are shown in Figure \ref{fig:mod2b}. Considering the left part of the plot, the cross-sectional effect is visualized along the baseline age axis, with a trajectory similar to the lifespan hippocampal volume shown in Figure \ref{fig:characteristic_curves}. The longitudinal effect, plotted along the time axis, is positive for low baseline ages and negative for higher baseline ages. The tensor interaction term plotted in the right part of Figure \ref{fig:mod2b} shows that the effect of time is positive in the youngest participants, quite flat in adults, and negative in the oldest participants. Note that the left and right plots in Figure \ref{fig:mod2b} are not comparable. Since the right plot is a pure interaction term, the direction of the estimated total longitudinal effect cannot be evaluated based on Figure \ref{fig:mod2b} (right) alone, but also needs to take the main effect of time into account.

\paragraph{Comparison of model fits}

Figure \ref{fig:cross_long_effects_hippocampus_example} shows estimated cross-sectional and longitudinal effects from the five models estimated in this section. The cross-sectional effects are estimated for 1st January 2010, and are all quite similar. Model (1a), estimated with only baseline measurements, indicates a less steep growth during childhood, and also exhibits some wiggliness between the age of 20 and the age of 50. The longitudinal effects are estimated 15 years ahead from baseline ages of 10, 30, 50, and 70 years. Models (1a) and (1b) do not distinguish longitudinal and cross-sectional effects, and hence have identical estimates in both plots. The estimates from model (2b) are quite different from those of the other models, except for a baseline age of 70. Given the simulation results of Section \ref{sec:simulations}, we suspect that the estimates from (2b) are not accurate. The longitudinal estimates of models (3a) and (3b) are close to the estimates of model (1b), again suggesting that the cohort effects in these data are moderate.

\paragraph{Estimating age at maximum volume}

\begin{figure}
\centering
\includegraphics{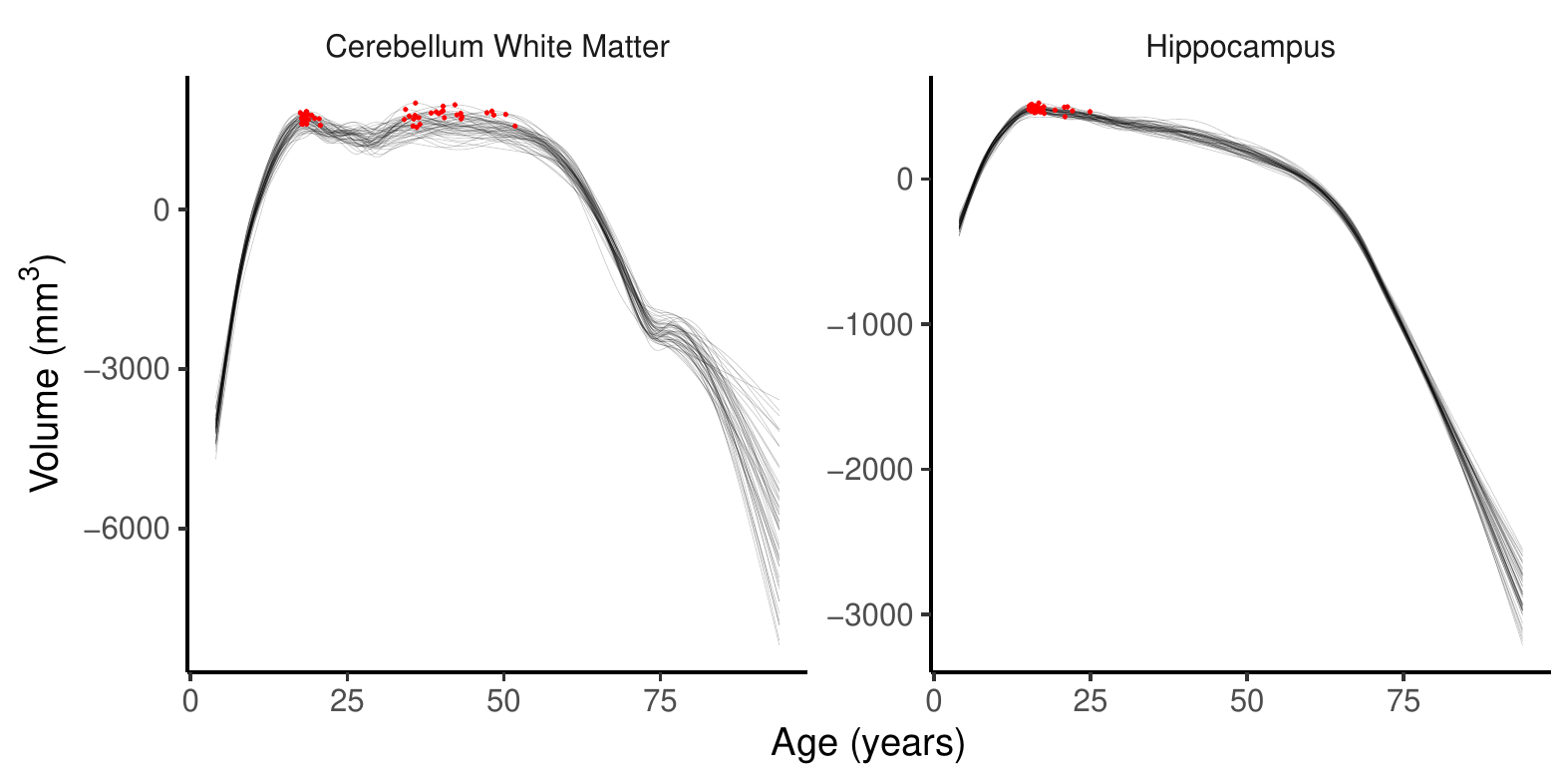}
\caption{\textbf{Posterior samples}. The plots show 50 samples from the posterior distributions of curves for lifespan volumes of cerebellum white matter and hippocampus. Red dots indicate the maximum of each curve.}
\label{fig:posterior_samples}
\end{figure}

A question of interest when estimating lifespan curves, is the age at which critical points occur, e.g. the age of maximum volume, maximum growth, or maximum decline. Point estimates of such critical ages can be read directly from the fits, if necessary after computing derivatives, but an assessment of their statistical uncertainty is not directly available. A Bayesian view of the smoothing introduced by \citet{Kimeldorf1970} lets us achieve this. Letting $\hat{\beta}$ denote the estimated regression parameters, including spline weights, and $\hat{\Sigma}_{\beta}$ their covariance matrix, the posterior distribution of the true coefficients $\beta$ is now a normal distribution with mean $\hat{\beta}$ and covariance $\hat{\Sigma}_{\beta}$, $\beta | y \sim N(\hat{\beta}, \hat{\Sigma}_{\beta})$ \citep[Ch. 6.10]{Wood2017}. By sampling from this posterior distribution we can make confidence statements about any quantity derived from the smooth functions. As an example, Figure \ref{fig:posterior_samples} shows 50 samples from the posterior distribution of volume curves for cerebellum white matter and hippocampus, with the maximum of each marked with a red dot. Even from these small samples it is evident that there is high uncertainty about the age at which cerebellum white matter volume is maximal, while there is less uncertainty about hippocampal volume.

\begin{figure}
\centering
\includegraphics{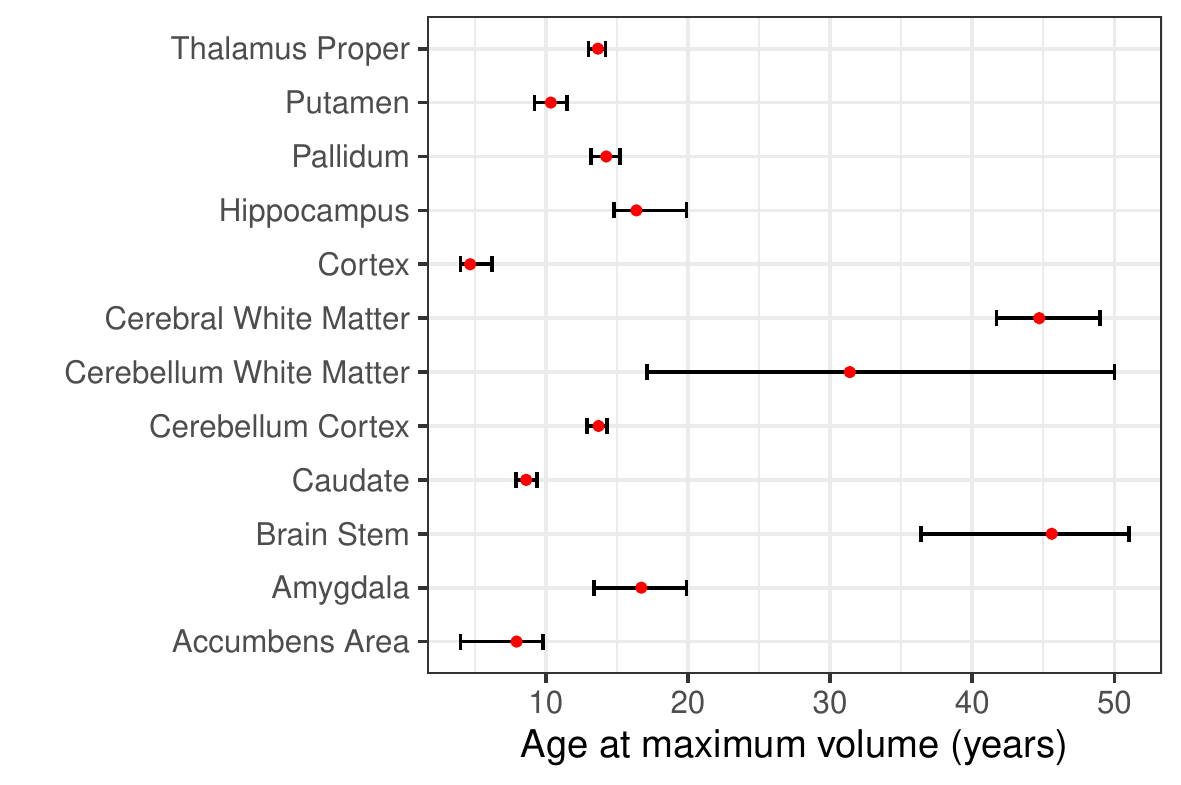}
\caption{\textbf{Age at maximum volume}. 95 \% highest posterior density intervals for the age at maximum volume of 12 brain regions. Red dots show posterior means.}
\label{fig:hdi_age}
\end{figure}

By sampling 20,000 curves and locating the age at maximum for each, we obtained posterior distributions of the age at maximum for each region. Figure \ref{fig:hdi_age} shows 95 \% highest posterior density intervals computed from the posterior distributions for all 12 regions, using the \verb!HDInterval! package \citep{Meredith2019}. The plot shows that the uncertainty about the location of the maximum is highly variable between regions, with very narrow intervals for, e.g. caudate, cerebellum cortex, and thalamus proper, and wide intervals and high uncertainty, e.g. for the brain stem and cerebellum white matter.

\subsubsection{Interaction effects on lifespan volume trajectories}
\label{sec:interactions}

We now demonstrate how factor-smooth interactions can be used to study how lifespan brain volumes are affected by a categorical variable. As an example application we consider the apolipoprotein E (APOE) $\epsilon$4 allele, which is a known risk factor for Alzheimer's disease \citep{Corder1993,Genin2011}, and study how lifespan trajectories of cerebellum cortex volume differ between carriers of  zero, one, or two APOE $\epsilon$4 alleles. Similar models were used by \citet{Walhovd2019}, who studied the impact of the APOE $\epsilon$4 allele on lifespan hippocampal volume.

The data are still contained in a dataframe named \verb!dat!, with identical structure to the data used in Section \ref{sec:LifespanHippovol}, except that the variable representing hippocampal volume now is replaced by the variable \verb!Cerebellum! representing cerebellum cortex volume in mm$^3$. In addition, a new variable \verb!Gene_APOEnE4! represents the total number of APOE $\epsilon$4 alleles. After excluding participants without information about APOE status, \verb!dat! contained 2,707 observations of 1,139 participants. Of these, 764 (1,838 observations) had zero alleles, 341 (789 observations) had 1 allele, and 34 (80 observations) had two alleles.

\paragraph{Factor smooths}

In order to estimate the interaction effects, the variable \verb!Gene_APOEnE4! needs to be coded as an ordered factor. This is done with the following code.

\begin{knitrout}
\definecolor{shadecolor}{rgb}{0.969, 0.969, 0.969}\color{fgcolor}\begin{kframe}
\begin{alltt}
\hlstd{dat}\hlopt{$}\hlstd{Gene_APOEnE4} \hlkwb{<-} \hlkwd{ordered}\hlstd{(dat}\hlopt{$}\hlstd{Gene_APOEnE4)}
\hlcom{# Print the levels of the ordered factor}
\hlkwd{levels}\hlstd{(dat}\hlopt{$}\hlstd{Gene_APOEnE4)}
\end{alltt}
\begin{verbatim}
## [1] "0" "1" "2"
\end{verbatim}
\end{kframe}
\end{knitrout}

A factor-smooth interaction is defined by \verb!s(Age, by = Gene_APOEnE4, k = 10, bs = "cr")!. For an ordered factor variable with $L$ levels, this term creates $L-1$ smooth functions, each representing the difference between the trajectory associated with the $l$th level ($l=2,\dots,L$) and the trajectory associated with the baseline level $l=1$. The difference between trajectories does not include pure offset effects, and hence the main effect of the ordered factor must be added. In this case, two smooth factor interaction terms are created, associated with 1 or 2 APOE $\epsilon$4 alleles. In contrast, if \verb!Gene_APOEnE4! was a numeric variable, \verb!gamm4()! would estimate a varying-coefficient term as used in Section \ref{sec:cohort_effects_varcoef} treating the number of alleles as a continuous variable, and if \verb!Gene_APOEnE4! was a factor variable a single smooth term would be independently estimated for each of the three factor levels and the main effect term \verb!s(Age, k = 10, bs = "cr")! would have to be omitted for the model to be identified.

\begin{knitrout}
\definecolor{shadecolor}{rgb}{0.969, 0.969, 0.969}\color{fgcolor}\begin{kframe}
\begin{alltt}
\hlstd{mod} \hlkwb{<-} \hlkwd{gamm4}\hlstd{(Cerebellum} \hlopt{~} \hlkwd{s}\hlstd{(Age,} \hlkwc{k} \hlstd{=} \hlnum{10}\hlstd{,} \hlkwc{bs} \hlstd{=} \hlstr{"cr"}\hlstd{)} \hlopt{+}
               \hlkwd{s}\hlstd{(Age,} \hlkwc{by} \hlstd{= Gene_APOEnE4,} \hlkwc{k} \hlstd{=} \hlnum{10}\hlstd{,} \hlkwc{bs} \hlstd{=} \hlstr{"cr"}\hlstd{)} \hlopt{+}
               \hlstd{Gene_APOEnE4} \hlopt{+} \hlstd{ICV_z} \hlopt{+} \hlstd{Sex} \hlopt{+} \hlstd{Scanner,} \hlkwc{data} \hlstd{= dat,} \hlkwc{random} \hlstd{=} \hlopt{~}\hlstd{(}\hlnum{1}\hlopt{|}\hlstd{ID))}
\end{alltt}
\end{kframe}
\end{knitrout}

Again, the matrix \verb!s.table! returned by \verb!summary()! can be used to study the smooth terms. The first term represents the trajectory for non-carriers, and the two terms starting with \verb!s(Age):Gene_APOEnE4! represent the difference between trajectories of carriers of one or two alleles to carriers of zero alleles, respectively. From the $p$-values for the interaction terms, it is clear that there is no evidence that the shape of the lifespan cerebellum white matter volume depends on APOE $\epsilon$4 status. 

\begin{knitrout}
\definecolor{shadecolor}{rgb}{0.969, 0.969, 0.969}\color{fgcolor}\begin{kframe}
\begin{alltt}
\hlkwd{summary}\hlstd{(mod}\hlopt{$}\hlstd{gam)}\hlopt{$}\hlstd{s.table}
\end{alltt}
\begin{verbatim}
##                        edf Ref.df       F p-value
## s(Age)               8.436  8.436 103.456   0.000
## s(Age):Gene_APOEnE41 1.000  1.000   0.424   0.515
## s(Age):Gene_APOEnE42 2.446  2.446   2.196   0.187
\end{verbatim}
\end{kframe}
\end{knitrout}

The main effects of APOE $\epsilon$4 status can be extracted from \verb!p.table! returned by \verb!summary()!. We use R's \verb!grep()! function to retain only terms containing the pattern "Gene\_APOE". The estimates are less than one standard error from zero and not significant, indicating that there is no evidence for an offset effect of APOE $\epsilon$4 status on cerebellum cortex volume. The suffixes \verb!.L! ('linear') and \verb!.Q! ('quadratic') are a consequence of how \verb!R! treats ordered factors, and represent the offset effect of having one or two alleles, respectively, relative to having zero alleles.

\begin{knitrout}
\definecolor{shadecolor}{rgb}{0.969, 0.969, 0.969}\color{fgcolor}\begin{kframe}
\begin{alltt}
\hlstd{ptab} \hlkwb{<-} \hlkwd{summary}\hlstd{(mod}\hlopt{$}\hlstd{gam)}\hlopt{$}\hlstd{p.table}
\hlcom{# Extract terms containing "Gene_APOE"}
\hlstd{ptab[}\hlkwd{grep}\hlstd{(}\hlstr{"Gene_APOE"}\hlstd{,} \hlkwd{rownames}\hlstd{(ptab)), ]}
\end{alltt}
\begin{verbatim}
##                Estimate Std. Error t value Pr(>|t|)
## Gene_APOEnE4.L   -766.3     1103.5 -0.6944   0.4875
## Gene_APOEnE4.Q   -509.8      754.3 -0.6758   0.4992
\end{verbatim}
\end{kframe}
\end{knitrout}

\paragraph{Prediction from GAMMs}

Creating predictions from GAMMs aids interpretation of the estimated effects, and we illustrate it here by comparing the estimated lifespan cerebellum cortex volumes for participants with zero, one, or two APOE $\epsilon$4 alleles. First, a grid over which to compute the predictions is created. Using \verb!expand.grid()!, all combinations of ages between 4 and 94 years with a spacing of 0.1 years, number of APOE $\epsilon$4 alleles, and sexes are generated. The \verb!predict()! function requires all variables in the model to be defined, and we hence set \verb!ICV_z! equal to the sample mean and \verb!Scanner! arbitrarily to \verb!"ousAvanto"!, which is one of the scanners used in the LCBC data. Other values of \verb!ICV_z! and \verb!Scanner! would shift the resulting curves vertically, but the interpretation would not change.

\begin{knitrout}
\definecolor{shadecolor}{rgb}{0.969, 0.969, 0.969}\color{fgcolor}\begin{kframe}
\begin{alltt}
\hlcom{# Create grid with all combinations of ages and APOE e4 alleles}
\hlstd{grid} \hlkwb{<-} \hlkwd{expand.grid}\hlstd{(}
  \hlkwc{Age} \hlstd{=} \hlkwd{seq}\hlstd{(}\hlkwc{from} \hlstd{=} \hlnum{4}\hlstd{,} \hlkwc{to} \hlstd{=} \hlnum{94}\hlstd{,} \hlkwc{by} \hlstd{=} \hlnum{.1}\hlstd{),} \hlkwc{Gene_APOEnE4} \hlstd{=} \hlkwd{ordered}\hlstd{(}\hlnum{0}\hlopt{:}\hlnum{2}\hlstd{),}
  \hlkwc{Sex} \hlstd{=} \hlkwd{factor}\hlstd{(}\hlkwd{c}\hlstd{(}\hlstr{"Female"}\hlstd{,} \hlstr{"Male"}\hlstd{)),} \hlkwc{ICV_z} \hlstd{=} \hlnum{0}\hlstd{,} \hlkwc{Scanner} \hlstd{=} \hlstr{"ousAvanto"}\hlstd{)}
\end{alltt}
\end{kframe}
\end{knitrout}

Next, predictions are computed at all values of the grid and \verb!ggplot2! \citep{Wickham2016} is used to plot the predicted values.

\begin{knitrout}
\definecolor{shadecolor}{rgb}{0.969, 0.969, 0.969}\color{fgcolor}\begin{kframe}
\begin{alltt}
\hlstd{grid}\hlopt{$}\hlstd{fit} \hlkwb{<-} \hlkwd{predict}\hlstd{(mod}\hlopt{$}\hlstd{gam,} \hlkwc{newdata} \hlstd{= grid)} \hlcom{# Compute fit on grid}

\hlkwd{library}\hlstd{(ggplot2)} \hlcom{# Plot with grouping by sex and number of alleles}
\hlkwd{ggplot}\hlstd{(grid,} \hlkwd{aes}\hlstd{(}\hlkwc{x} \hlstd{= Age,} \hlkwc{y} \hlstd{= fit,} \hlkwc{group} \hlstd{=} \hlkwd{interaction}\hlstd{(Gene_APOEnE4, Sex),}
                 \hlkwc{color} \hlstd{= Gene_APOEnE4,} \hlkwc{linetype} \hlstd{= Sex))} \hlopt{+} \hlkwd{geom_line}\hlstd{()}
\end{alltt}
\end{kframe}
\end{knitrout}

\begin{figure}
\centering
\includegraphics{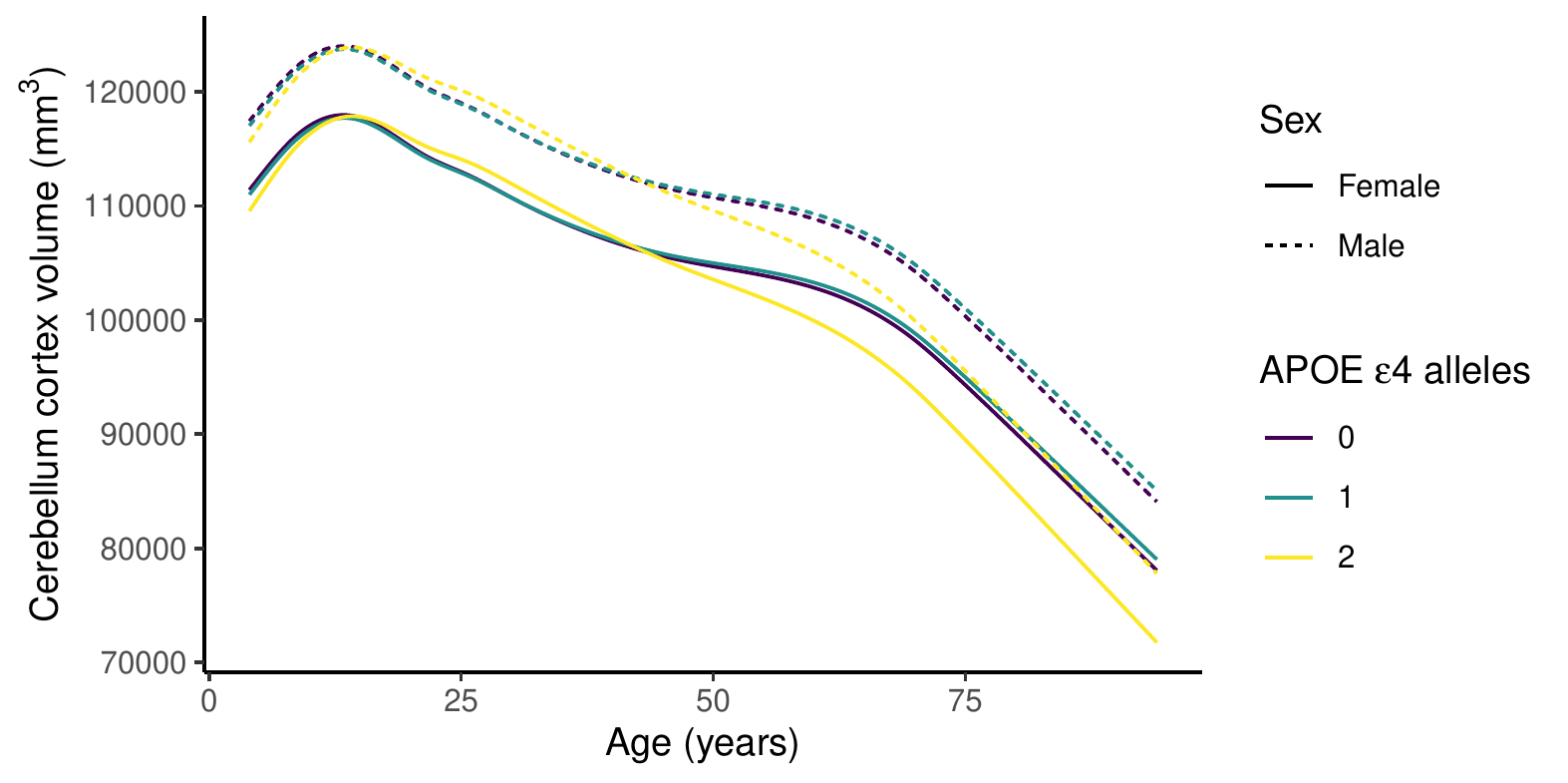}
\cprotect\caption{\textbf{Factor smooth interactions}. Estimated lifespan trajectories of cerebellum cortex volume for males and females with 0, 1, or 2 APOE $\epsilon$4 alleles.}
\label{fig:APOE_interaction}
\end{figure}

A slightly modified version of the resulting plot is shown in Figure \ref{fig:APOE_interaction}. Note that since \verb!Sex! is a parametric term, it merely shifts the curves vertically, without changing their shapes. In this example, none of the interaction effects were significant, but the plots still show how smooth interaction terms create different functional shapes depending on the number of APOE $\epsilon$4 alleles.

\section{Discussion}
\label{sec:discussion}

This paper has highlighted that GAMMs are well-suited for estimating lifespan brain trajectories. However, the issue of potential cohort effects requires careful consideration, and direct translation of LMM formulations used for separating longitudinal and cross-sectional effects has potential pitfalls. In Section \ref{sec:GAMMlong} we defined the "age model" \eqref{eq:GAMM1} which ignores cohort effects, the "age-time model" \eqref{eq:GAMM2} which is a direct extension of LMMs commonly used to separate longitudinal and cross-sectional effects, and the alternative "age-cohort model" \eqref{eq:GAMM3} which includes participant birth date as a model term. These models' abilities to accurately estimate longitudinal and cross-sectional effects were compared in realistic simulation experiments reported in Section \ref{sec:simulations}. Not surprisingly, in the absence of cohort effects the age model was most accurate, with the version using longitudinal data performing better than the equivalent model with only cross-sectional data. With cohort effects, on the other hand, the age cohort model was most accurate. More importantly, the age-time model  -- which may be seen as a "classic" model used to separate longitudinal and cohort effects -- consistently performed worse than the age-cohort model, and as shown in Table \ref{tab:simres} it had both higher bias and variance across the simulated samples. As also suggested by Figures \ref{fig:simres_long_rmse_summary_varying_date_no_cohort} and \ref{fig:simres_long_rmse_summary_varying_date_cohort_interactions}, the age-time model is not able to estimate nonlinear longitudinal effects beyond short follow-up intervals. On the other hand, in the special case of linear longitudinal effects, from which the age-time model originates, longitudinal effects for any time after baseline can in principle be accurately estimated with follow-up intervals of arbitrary length. Interpretation of the terms in the age-time model as longitudinal and cross-sectional effects also requires that all participants have equal dates of initial measurement. However, simulations reported in Supplementary Section S2.4 suggest that varying baseline dates have a very small effect on the accuracy of the age-time model in the settings considered here.

While limitations will vary with regard to study specific characteristics, we find it important to emphasize, in light of the present findings, that age-time models will never be able to make accurate estimates of lifespan trajectories, or even trajectories for any substantial part of the lifespan, unless a majority of participants have been followed over the whole interval of interest. Currently available human cohorts with longitudinal imaging data do not span the desired intervals. Furthermore, reaching acceptable power is impossible if dates of initial measurement are to be contained within a small fraction of time. A look at some of the most powerful and impressive combined cross-sectional and longitudinal studies of brain changes with age, suggests that fractional follow-up interval and range of variation in initial measurement dates realistically need to be accommodated in all statistical models. For instance, even the ABCD study \citep{Casey2018} which utilizes numerous scan sites to track development in thousands of children at very similar age, need to allow for some variation in initial date of measurement, and follow-up intervals are so far limited to a couple of years. While there luckily are major and most impressive studies that contain information on participant samples for many decades, such as the Whitehall study \citep{Filippini2014}, the Baltimore Longitudinal Study of Aging \citep{Tian2015}, the Betula \citep{Gorbach2017}, or the Lothian Birth Cohort study \citep{Cox2018} these still await longitudinal imaging data \citep{Filippini2014}, or typically have MRI data only for a small fraction of the time, less than a decade \citep{Cox2018,Gorbach2017,Tian2015}, sometimes with scan waves being completed across several years \citep{Gorbach2017}. We note this not as a critique of any study, but as a reminder that statistical models at the very least need to accomodate the realistic situation for the best possible data.

\subsection{When is a single measurement per participant sufficient?}
The dangers of using cross-sectional data have repeatedly been pointed out in the quantitative psychology literature. For example, mediation analysis using purely cross-sectional data is likely to lead to biased and misleading estimates under realistic conditions \citep{Cole2003,Lindenberger2011,Lindenberger1998}. In mediation analysis, the goal is to understand the causal paths through which one or more variables $x$ influence an outcome $y$, directly or through one or more mediating variables $m$. Since a cause precedes its effects, carefully designed longitudinal data collection as well as models capable of utilizing this information are then necessary \citep{Collins1998}. Longitudinal data is also required to understand how within-individual change differs from between-individual change and how within-individual change is correlated across multiple processes \citep{Lindenberger2011,Molenaar2004}. Traditionally, such studies have been conducted by following a group of participants of similar age over a number of waves, e.g. \citet{Cox2020,Raz2005,Raz2010}.

While the above mentioned cautions about use of cross-sectional data are completely justified, they do not necessarily extrapolate to estimation of lifespan trajectories. If the goal is to estimate the population effect of aging on the volume of one or more brain regions, potentially including interaction effects of static trait variables like genetic variations or education level (after completed education), a single measurement per participant may be sufficient. One example is when the strong assumption of no cohort effects is made. If it holds, cross-sectional and longitudinal effects are equal, and both can be accurately estimated by the age model using purely cross-sectional data. However, with sufficient variation in baseline dates, the age-cohort model is in principle able to estimate longitudinal and cross-sectional effects using a single measurement per participant. This approach has been used in studies of cognitive aging, which allows estimation of aging and cohort effects without the risk of confounding by retest effects \citep{Horn1976,Salthouse2013,Salthouse2014,Salthouse2019,Schaie1973}. In practice, however, we have experienced that cohort-age models become more stable and accurate with longitudinal data. In particular, both the additional variation provided by repeated measurements with heterogeneous follow-up intervals and the correlation between repeated measurements of the same participant likely contribute to better separation of age effects and cohort effects. Furthermore, as shown in Section \ref{sec:LifespanHippovol}, a GAMM using longitudinal data also estimates the between-individual variation and the within-individual variation, quantifying the extent to which differences between participants are due to systematic variation and noise, respectively.

\subsection{Limitations and future directions}
The GAMMs studied in this paper also have some limitations. The age-cohort model is not identified if all participants have been measured at the exact same dates, since age and cohort then are perfectly collinear. This is also true with longitudinal data, and emphasizes the fact that these models are developed for heterogeneous data, typically combined from multiple studies.

There is also a need for methodological development related to estimation of correlated change between regions. In principle, this could be done by fitting GAMMs separately for each region, and using the correlation of random slopes across regions as an estimate of correlated change. A more principled approach is offered by joint modeling frameworks for LMMs \citep{Fieuws2004,Fieuws2006}, which in this case would amount to fitting a single hierarchical GAMM \citep{Pedersen2019} for the lifespan trajectories of multiple regions, with interaction terms distinguishing trajectories for each region and random effect structures modeling the within-individual level and change correlation between region trajectories. However, the fact that the extent of correlated change between any pair of brain regions is likely to vary across the lifespan would also need to be taken into account, e.g. by modeling correlations as functions of age. Combined with the need for three or more timepoints to accurately estimate random slopes, it may currently be challenging to obtain a sufficient amount of longitudinal data for fitting such models.

A limitation of GAMMs is that the parameters of the estimated smooth terms typically are not interpretable, and quantities such as rates of change, initial levels, and final asymptotic levels have to be inferred from the estimated functions, e.g. as demonstrated for age at maximum volume in Figures \ref{fig:posterior_samples} and \ref{fig:hdi_age} in Section \ref{sec:LifespanHippovol}. When prior knowledge about the functional form of the phenomenon under study is available, nonlinear mixed models are an attractive alternative, offering directly interpretable parameters of substantive interest \citep{Davidian2009,Lindstrom1990,Ram2007}. Applications include modeling of learning curves \citep{Cudeck1996,Ghisletta2010}, modeling the effect of preschool instructions on academic achievement using sigmoidal functions \citep{Grimm2009}, and modeling of rate of change and acceleration in a lexical retrieval task \citep{Grimm2013}.

While we have considered GAMMs for estimating how time-invariant variables interact with lifespan trajectories in Section \ref{sec:interactions}, interaction with time-dependent variables may also be of interest. Although time-dependent interaction variables can be used within the framework considered here, the interpretation of the estimated effects becomes more challenging. If only the value of the time-dependent variable at the given timepoint  affects the outcome, the effect can be interpreted in exactly the same way as for a time-invariant variable \citep[Ch. 13.5]{Fitzmaurice2011}. In many applications, however, it may be more plausible to assume that also the variable's change since the previous timepoint contains relevant information, and in this case the models used in Section \ref{sec:interactions} will have biased estimates. Continuous-time SEMs \citep{Driver2018,Oud2000} may be useful for this purpose, as they allow regressing a time-dependent process (the outcome of interest) on the value of another time-dependent process (the interaction variable) at earlier times, without the restrictive assumption of equally spaced time intervals imposed by ordinary SEMs. However, estimation of nonlinear smooth functions within continuous-time SEMs has not been reported in the literature, and is likely to be both computationally challenging and require large longitudinal datasets.

\section{Conclusion}
\label{sec:conclusion}

GAMMs are attractive tools for estimating lifespan brain trajectories, which flexibly handle the nonlinear effects and variable follow-up intervals and measurement dates characteristic of lifespan data. If cohort effects are negligible, age models on the form \eqref{eq:GAMM1} yield the most accurate estimates, and in this case a single measurement per participant may even be sufficient. More realistically, cohort effects are likely to be present, and in this case the age-cohort model \eqref{eq:GAMM3} which directly models the effect of birth cohort is able to accurately estimate longitudinal and cross-sectional effects. On the other hand, the age-time model \eqref{eq:GAMM2} which separates the effect of age into a baseline term and a time term as is common with LMMs, yields poor estimates of longitudinal effects. With sufficient variation of measurement dates and follow-up intervals, we thus recommend the age-cohort model for estimating lifespan brain trajectories. On the other hand, for time-structured data containing little variation in measurement dates, the age-cohort model is not identified and the age-time model seems to be the best option, with the caveat that estimated longitudinal effects may not be reliable for times larger than the average follow-up interval, as will also be apparent from the confidence intervals. 

The \verb!R! packages \verb!mgcv! and \verb!gamm4! provide efficient software for fitting GAMMs, and are complemented by additional packages enabling easy visualization and interpretation of summary statistics.

\section*{Declaration of Competing Interest}
The authors report no competing interests.

\section*{Data and Code Availability Statement}
Supplementary Material for this arXiv preprint is available at https://osf.io/xnuhz/.

Code for running simulation experiments is available in Supplementary Section S4 and at the Open Science Framework, https://osf.io/xnuhz/. The LCBC data are available from the corresponding author on reasonable request, given appropriate ethical and data protection approvals.

\section*{Ethics Statement}
Collection of the LCBC was approved by a Norwegian Regional Committee for Medical and Health Research Ethics South. Written informed consent was obtained from all participants of age at majority, and the parents or guardian for children below this age.

\section*{Acknowledgement}
The authors thank Andreas Brandmaier, Paolo Ghisletta, and Magne Thoresen for helpful discussions. The data collection was supported by the European Research Council under grant agreements 283634, 725025 (to A.M.F.) and 313440 (to K.B.W.), the Norwegian Research Council (to A.M.F., K.B.W.), The National Association for Public Health's dementia research program, Norway (to A.M.F) and center support from the University of Oslo.

\bibliography{references}

\end{document}

%% file: tables/simres_varying_date.txt
\begin{tabular}[t]{lllllllllll}
\toprule
\multicolumn{2}{c}{ } & \multicolumn{3}{c}{No cohort effect} & \multicolumn{3}{c}{Age-indep. cohort effect} & \multicolumn{3}{c}{Cohort-age interaction} \\
\cmidrule(l{3pt}r{3pt}){3-5} \cmidrule(l{3pt}r{3pt}){6-8} \cmidrule(l{3pt}r{3pt}){9-11}
Region & Model & RMSE & Bias & $\sqrt{\text{Var.}}$ & RMSE & Bias & $\sqrt{\text{Var.}}$ & RMSE & Bias & $\sqrt{\text{Var.}}$\\
\midrule
Cerebral White Matter & (1a) & 2575. & 933.2 & 2401. & 4454. & 3758. & 2392. & 5071. & 4435. & 2460.\\
Cerebral White Matter & (1b) & 1995. & 419.5 & 1952. & 3801. & 3259. & 1957. & 4487. & 4028. & 1979.\\
Cerebral White Matter & (2a) & 3639. & 2493. & 2652. & 3789. & 2710. & 2650. & 3842. & 2776. & 2658.\\
Cerebral White Matter & (2b) & 3281. & 1536. & 2901. & 3646. & 2251. & 2870. & 3726. & 2302. & 2931.\\
Cerebral White Matter & (3a) & 2376. & 407.0 & 2342. & 2339. & 418.3 & 2303. & 3129. & 2091. & 2329.\\
Cerebral White Matter & (3b) & 2885. & 641.9 & 2814. & 2858. & 662.2 & 2781. & 2860. & 666.7 & 2782.\\
\midrule
Cortex & (1a) & 4843. & 3446. & 3405. & 6144. & 5109. & 3414. & 7408. & 6598. & 3369.\\
Cortex & (1b) & 3069. & 1128. & 2856. & 5044. & 4164. & 2848. & 6205. & 5509. & 2858.\\
Cortex & (2a) & 5907. & 4580. & 3734. & 6060. & 4761. & 3750. & 6263. & 4988. & 3789.\\
Cortex & (2b) & 4877. & 2330. & 4287. & 5627. & 3593. & 4332. & 5658. & 3580. & 4384.\\
Cortex & (3a) & 3438. & 1160. & 3238. & 3410. & 1093. & 3232. & 4697. & 3351. & 3293.\\
Cortex & (3b) & 3966. & 1215. & 3777. & 4045. & 1120. & 3889. & 4052. & 1129. & 3893.\\
\midrule
Hippocampus & (1a) & 45.01 & 19.20 & 40.73 & 73.93 & 61.70 & 40.74 & 83.46 & 72.65 & 41.11\\
Hippocampus & (1b) & 32.73 & 7.817 & 31.80 & 60.90 & 51.83 & 31.99 & 71.86 & 64.41 & 31.87\\
Hippocampus & (2a) & 77.75 & 64.06 & 44.09 & 80.04 & 66.48 & 44.60 & 80.91 & 67.50 & 44.64\\
Hippocampus & (2b) & 60.92 & 35.30 & 49.68 & 69.19 & 46.32 & 51.41 & 68.12 & 44.75 & 51.39\\
Hippocampus & (3a) & 37.55 & 7.463 & 36.82 & 38.00 & 6.732 & 37.41 & 49.72 & 33.12 & 37.10\\
Hippocampus & (3b) & 45.08 & 9.353 & 44.12 & 45.52 & 8.306 & 44.78 & 45.99 & 8.622 & 45.20\\
\bottomrule
\end{tabular}

%% file: manuscript.bbl
\begin{thebibliography}{87}
\providecommand{\natexlab}[1]{#1}
\providecommand{\url}[1]{\texttt{#1}}
\expandafter\ifx\csname urlstyle\endcsname\relax
  \providecommand{\doi}[1]{doi: #1}\else
  \providecommand{\doi}{doi: \begingroup \urlstyle{rm}\Url}\fi

\bibitem[Baltes(1968)]{Baltes1968}
Baltes, P.B.
\newblock Longitudinal and cross-sectional sequences in the study of age and
  generation effects.
\newblock \emph{Human Development}, 11\penalty0 (3):\penalty0 145--171, 1968.
\newblock \doi{10.1159/000270604}.
\newblock URL \url{https://doi.org/10.1159/000270604}.

\bibitem[Bates et~al.(2015)Bates, M{\"a}chler, Bolker, and Walker]{Bates2015}
Bates, Douglas; M{\"a}chler, Martin; Bolker, Ben, and Walker, Steve.
\newblock Fitting linear mixed-effects models using {lme4}.
\newblock \emph{Journal of Statistical Software}, 67\penalty0 (1):\penalty0
  1--48, 2015.
\newblock \doi{10.18637/jss.v067.i01}.
\newblock URL \url{https://doi.org/10.18637/jss.v067.i01}.

\bibitem[Bearden and Thompson(2017)]{Bearden2017}
Bearden, Carrie~E. and Thompson, Paul~M.
\newblock Emerging global initiatives in neurogenetics: The enhancing
  neuroimaging genetics through meta-analysis ({ENIGMA}) consortium.
\newblock \emph{Neuron}, 94\penalty0 (2):\penalty0 232--236, April 2017.
\newblock \doi{10.1016/j.neuron.2017.03.033}.
\newblock URL \url{https://doi.org/10.1016/j.neuron.2017.03.033}.

\bibitem[Brumback and Rice(1998)]{Brumback1998}
Brumback, Babette~A. and Rice, John~A.
\newblock Smoothing spline models for the analysis of nested and crossed
  samples of curves.
\newblock \emph{Journal of the American Statistical Association}, 93\penalty0
  (443):\penalty0 961--976, 1998.
\newblock \doi{10.1080/01621459.1998.10473755}.
\newblock URL \url{https://doi.org/10.1080/01621459.1998.10473755}.

\bibitem[Casey et~al.(2018)Casey, Cannonier, Conley, Cohen, Barch, Heitzeg,
  Soules, Teslovich, Dellarco, Garavan, Orr, Wager, Banich, Speer, Sutherland,
  Riedel, Dick, Bjork, Thomas, Chaarani, Mejia, Hagler, Cornejo, Sicat, Harms,
  Dosenbach, Rosenberg, Earl, Bartsch, Watts, Polimeni, Kuperman, Fair, and
  Dale]{Casey2018}
Casey, B.J.; Cannonier, Tariq; Conley, May~I.; Cohen, Alexandra~O.; Barch,
  Deanna~M.; Heitzeg, Mary~M.; Soules, Mary~E.; Teslovich, Theresa; Dellarco,
  Danielle~V.; Garavan, Hugh; Orr, Catherine~A.; Wager, Tor~D.; Banich,
  Marie~T.; Speer, Nicole~K.; Sutherland, Matthew~T.; Riedel, Michael~C.; Dick,
  Anthony~S.; Bjork, James~M.; Thomas, Kathleen~M.; Chaarani, Bader; Mejia,
  Margie~H.; Hagler, Donald~J.; Cornejo, M.~Daniela; Sicat, Chelsea~S.; Harms,
  Michael~P.; Dosenbach, Nico~U.F.; Rosenberg, Monica; Earl, Eric; Bartsch,
  Hauke; Watts, Richard; Polimeni, Jonathan~R.; Kuperman, Joshua~M.; Fair,
  Damien~A., and Dale, Anders~M.
\newblock The adolescent brain cognitive development ({ABCD}) study: Imaging
  acquisition across 21 sites.
\newblock \emph{Developmental Cognitive Neuroscience}, 32:\penalty0 43--54,
  August 2018.
\newblock \doi{10.1016/j.dcn.2018.03.001}.
\newblock URL \url{https://doi.org/10.1016/j.dcn.2018.03.001}.

\bibitem[Cole and Maxwell(2003)]{Cole2003}
Cole, David~A. and Maxwell, Scott~E.
\newblock Testing mediational models with longitudinal data: Questions and tips
  in the use of structural equation modeling.
\newblock \emph{Journal of Abnormal Psychology}, 112\penalty0 (4):\penalty0
  558--577, 2003.
\newblock \doi{10.1037/0021-843x.112.4.558}.
\newblock URL \url{https://doi.org/10.1037/0021-843x.112.4.558}.

\bibitem[Collins et~al.(1998)Collins, Graham, and Flaherty]{Collins1998}
Collins, Linda~M.; Graham, John~J., and Flaherty, Brian~P.
\newblock An alternative framework for defining mediation.
\newblock \emph{Multivariate Behavioral Research}, 33\penalty0 (2):\penalty0
  295--312, 1998.
\newblock \doi{10.1207/s15327906mbr3302_5}.
\newblock URL \url{https://doi.org/10.1207/s15327906mbr3302_5}.

\bibitem[Corder et~al.(1993)Corder, Saunders, Strittmatter, Schmechel, Gaskell,
  Small, Roses, Haines, and Pericak-Vance]{Corder1993}
Corder, E.; Saunders, A.; Strittmatter, W.; Schmechel, D.; Gaskell, P.; Small,
  G.; Roses, A.; Haines, J., and Pericak-Vance, M.
\newblock Gene dose of apolipoprotein e type 4 allele and the risk of
  alzheimer's disease in late onset families.
\newblock \emph{Science}, 261\penalty0 (5123):\penalty0 921--923, 1993.
\newblock \doi{10.1126/science.8346443}.
\newblock URL \url{https://doi.org/10.1126/science.8346443}.

\bibitem[Cox et~al.(2018)Cox, Allerhand, Ritchie, Maniega, Hern{\'{a}}ndez,
  Harris, Dickie, Anblagan, Aribisala, Morris, Sherwood, Abbott, Starr, Bastin,
  Wardlaw, and Deary]{Cox2018}
Cox, Simon~R.; Allerhand, Mike; Ritchie, Stuart~J.; Maniega,
  Susana~Mu{\~{n}}oz; Hern{\'{a}}ndez, Maria~Vald{\'{e}}s; Harris, Sarah~E.;
  Dickie, David~Alexander; Anblagan, Devasuda; Aribisala, Benjamin~S.; Morris,
  Zoe; Sherwood, Roy; Abbott, N.~Joan; Starr, John~M.; Bastin, Mark~E.;
  Wardlaw, Joanna~M., and Deary, Ian~J.
\newblock Longitudinal serum s100$\beta$ and brain aging in the {Lothian}
  {Birth} {Cohort} 1936.
\newblock \emph{Neurobiology of Aging}, 69:\penalty0 274--282, September 2018.
\newblock \doi{10.1016/j.neurobiolaging.2018.05.029}.
\newblock URL \url{https://doi.org/10.1016/j.neurobiolaging.2018.05.029}.

\bibitem[Cox et~al.(2020)Cox, Harris, Ritchie, Buchanan, Hern{\'a}ndez, Corley,
  Taylor, Madole, Harris, Whalley, McIntosh, Russ, Bastin, Wardlaw, Deary, and
  Tucker-Drob]{Cox2020}
Cox, SR; Harris, MA; Ritchie, SJ; Buchanan, CR; Hern{\'a}ndez,
  Maria~Vald{\'e}s; Corley, Janie; Taylor, Adele~M; Madole, JW; Harris, SE;
  Whalley, HC; McIntosh, AM; Russ, TC; Bastin, ME; Wardlaw, JM; Deary, IJ, and
  Tucker-Drob, EM.
\newblock Three major dimensions of human brain cortical ageing in relation to
  cognitive decline across the 8th decade of life.
\newblock \emph{bioRxiv}, 2020.
\newblock \doi{10.1101/2020.01.19.911420}.
\newblock URL
  \url{https://www.biorxiv.org/content/early/2020/02/16/2020.01.19.911420}.

\bibitem[Cudeck(1996)]{Cudeck1996}
Cudeck, Robert.
\newblock Mixed-effects models in the study of individual differences with
  repeated measures data.
\newblock \emph{Multivariate Behavioral Research}, 31\penalty0 (3):\penalty0
  371--403, July 1996.
\newblock \doi{10.1207/s15327906mbr3103_6}.
\newblock URL \url{https://doi.org/10.1207/s15327906mbr3103_6}.

\bibitem[Curran and Bauer(2011)]{Curran2011}
Curran, Patrick~J. and Bauer, Daniel~J.
\newblock The disaggregation of within-person and between-person effects in
  longitudinal models of change.
\newblock \emph{Annual Review of Psychology}, 62\penalty0 (1):\penalty0
  583--619, 2011.
\newblock \doi{10.1146/annurev.psych.093008.100356}.
\newblock URL \url{https://doi.org/10.1146/annurev.psych.093008.100356}.

\bibitem[Dale et~al.(1999)Dale, Fischl, and Sereno]{Dale1999}
Dale, Anders~M.; Fischl, Bruce, and Sereno, Martin~I.
\newblock Cortical surface-based analysis.
\newblock \emph{{NeuroImage}}, 9\penalty0 (2):\penalty0 179--194, 1999.
\newblock \doi{10.1006/nimg.1998.0395}.
\newblock URL \url{https://doi.org/10.1006/nimg.1998.0395}.

\bibitem[Davidian(2009)]{Davidian2009}
Davidian, M.
\newblock Non-linear mixed effects models.
\newblock In Fitzmaurice, G.; Davidian, M.; Verbeke, G., and Molenberghs, G.,
  editors, \emph{Longitudinal Data Analysis}, pages 291--316. Chapman and Hall,
  London, U.K., 2009.

\bibitem[Diggle et~al.(2002)Diggle, Heagerty, Liang, and Zeger]{Diggle2002}
Diggle, Peter; Heagerty, Patrick; Liang, Kung-Yee, and Zeger, Scott.
\newblock \emph{Analysis of Longitudinal Data 2nd Edition}.
\newblock Oxford University Press, Oxford, U.K., 2002.

\bibitem[Driver and Voelkle(2018)]{Driver2018}
Driver, Charles~C. and Voelkle, Manuel~C.
\newblock Hierarchical bayesian continuous time dynamic modeling.
\newblock \emph{Psychological Methods}, 23\penalty0 (4):\penalty0 774--799,
  2018.
\newblock \doi{10.1037/met0000168}.
\newblock URL \url{https://doi.org/10.1037/met0000168}.

\bibitem[Durb{\'{a}}n et~al.(2005)Durb{\'{a}}n, Harezlak, Wand, and
  Carroll]{Durban2005}
Durb{\'{a}}n, M.; Harezlak, J.; Wand, M.~P., and Carroll, R.~J.
\newblock Simple fitting of subject-specific curves for longitudinal data.
\newblock \emph{Statistics in Medicine}, 24\penalty0 (8):\penalty0 1153--1167,
  2005.
\newblock \doi{10.1002/sim.1991}.
\newblock URL \url{https://doi.org/10.1002/sim.1991}.

\bibitem[Edwards et~al.(2005)Edwards, Stewart, MacDougall, and
  Helms]{Edwards2005}
Edwards, Lloyd~J.; Stewart, Paul~W.; MacDougall, James~E., and Helms, Ronald~W.
\newblock A method for fitting regression splines with varying polynomial order
  in the linear mixed model.
\newblock \emph{Statistics in Medicine}, 25\penalty0 (3):\penalty0 513--527,
  2005.
\newblock \doi{10.1002/sim.2232}.
\newblock URL \url{https://doi.org/10.1002/sim.2232}.

\bibitem[Fieuws and Verbeke(2004)]{Fieuws2004}
Fieuws, Steffen and Verbeke, Geert.
\newblock Joint modelling of multivariate longitudinal profiles: pitfalls of
  the random-effects approach.
\newblock \emph{Statistics in Medicine}, 23\penalty0 (20):\penalty0 3093--3104,
  2004.
\newblock \doi{10.1002/sim.1885}.
\newblock URL \url{https://doi.org/10.1002/sim.1885}.

\bibitem[Fieuws and Verbeke(2006)]{Fieuws2006}
Fieuws, Steffen and Verbeke, Geert.
\newblock Pairwise fitting of mixed models for the joint modeling of
  multivariate longitudinal profiles.
\newblock \emph{Biometrics}, 62\penalty0 (2):\penalty0 424--431, 2006.
\newblock \doi{10.1111/j.1541-0420.2006.00507.x}.
\newblock URL \url{https://doi.org/10.1111/j.1541-0420.2006.00507.x}.

\bibitem[Filippini et~al.(2014)Filippini, Zsoldos, Haapakoski, Sexton, Mahmood,
  Allan, Topiwala, Valkanova, Brunner, Shipley, Auerbach, Moeller,
  U{\u{g}}urbil, Xu, Yacoub, Andersson, Bijsterbosch, Clare, Griffanti, Hess,
  Jenkinson, Miller, Salimi-Khorshidi, Sotiropoulos, Voets, Smith, Geddes,
  Singh-Manoux, Mackay, Kivim\"{a}ki, and Ebmeier]{Filippini2014}
Filippini, Nicola; Zsoldos, Enik{\H{o}}; Haapakoski, Rita; Sexton, Claire~E;
  Mahmood, Abda; Allan, Charlotte~L; Topiwala, Anya; Valkanova, Vyara; Brunner,
  Eric~J; Shipley, Martin~J; Auerbach, Edward; Moeller, Steen; U{\u{g}}urbil,
  K{\^{a}}mil; Xu, Junqian; Yacoub, Essa; Andersson, Jesper; Bijsterbosch,
  Janine; Clare, Stuart; Griffanti, Ludovica; Hess, Aaron~T; Jenkinson, Mark;
  Miller, Karla~L; Salimi-Khorshidi, Gholamreza; Sotiropoulos, Stamatios~N;
  Voets, Natalie~L; Smith, Stephen~M; Geddes, John~R; Singh-Manoux, Archana;
  Mackay, Clare~E; Kivim\"{a}ki, Mika, and Ebmeier, Klaus~P.
\newblock Study protocol: the whitehall {II} imaging sub-study.
\newblock \emph{{BMC} Psychiatry}, 14\penalty0 (1), 2014.
\newblock \doi{10.1186/1471-244x-14-159}.
\newblock URL \url{https://doi.org/10.1186/1471-244x-14-159}.

\bibitem[Fischl et~al.(2002)Fischl, Salat, Busa, Albert, Dieterich, Haselgrove,
  van~der Kouwe, Killiany, Kennedy, Klaveness, Montillo, Makris, Rosen, and
  Dale]{Fischl2002}
Fischl, Bruce; Salat, David~H.; Busa, Evelina; Albert, Marilyn; Dieterich,
  Megan; Haselgrove, Christian; van~der Kouwe, Andre; Killiany, Ron; Kennedy,
  David; Klaveness, Shuna; Montillo, Albert; Makris, Nikos; Rosen, Bruce, and
  Dale, Anders~M.
\newblock Whole brain segmentation.
\newblock \emph{Neuron}, 33\penalty0 (3):\penalty0 341--355, 2002.
\newblock \doi{10.1016/s0896-6273(02)00569-x}.
\newblock URL \url{https://doi.org/10.1016/s0896-6273(02)00569-x}.

\bibitem[Fitzmaurice et~al.(2011)Fitzmaurice, Laird, and Ware]{Fitzmaurice2011}
Fitzmaurice, Garrett~M.; Laird, Nan~M., and Ware, James~H.
\newblock \emph{Applied Longitudinal Analysis 2nd Edition}.
\newblock John Wiley \& Sons, Inc., Hoboken, New Jersey, USA, 2011.

\bibitem[Fjell et~al.(2010)Fjell, Walhovd, Westlye, {\O}stby, Tamnes, Jernigan,
  Gamst, and Dale]{Fjell2010}
Fjell, Anders.~M.; Walhovd, Kristine~B.; Westlye, Lars~T.; {\O}stby, Ylva;
  Tamnes, Christian~K.; Jernigan, Terry~L.; Gamst, Anthony, and Dale, Anders~M.
\newblock When does brain aging accelerate? {D}angers of quadratic fits in
  cross-sectional studies.
\newblock \emph{NeuroImage}, 50\penalty0 (4):\penalty0 1376 -- 1383, 2010.
\newblock ISSN 1053-8119.
\newblock \doi{https://doi.org/10.1016/j.neuroimage.2010.01.061}.
\newblock URL
  \url{http://www.sciencedirect.com/science/article/pii/S1053811910000832}.

\bibitem[Fjell et~al.(2017)Fjell, Idland, Sala-Llonch, Watne, Borza,
  Br{\ae}khus, Lona, Zetterberg, Blennow, Wyller, and Walhovd]{Fjell2017}
Fjell, Anders~Martin; Idland, Ane-Victoria; Sala-Llonch, Roser; Watne,
  Leiv~Otto; Borza, Tom; Br{\ae}khus, Anne; Lona, Tarjei; Zetterberg, Henrik;
  Blennow, Kaj; Wyller, Torgeir~Bruun, and Walhovd, Kristine~Beate.
\newblock Neuroinflammation and tau interact with amyloid in predicting sleep
  problems in aging independently of atrophy.
\newblock \emph{Cerebral Cortex}, 28\penalty0 (8):\penalty0 2775--2785, 2017.
\newblock \doi{10.1093/cercor/bhx157}.
\newblock URL \url{https://doi.org/10.1093/cercor/bhx157}.

\bibitem[Genin et~al.(2011)Genin, Hannequin, Wallon, Sleegers, Hiltunen,
  Combarros, Bullido, Engelborghs, Deyn, Berr, Pasquier, Dubois, Tognoni,
  Fi{\'{e}}vet, Brouwers, Bettens, Arosio, Coto, Zompo, Mateo, Epelbaum,
  Frank-Garcia, Helisalmi, Porcellini, Pilotto, Forti, Ferri, Scarpini,
  Siciliano, Solfrizzi, Sorbi, Spalletta, Valdivieso, Veps\"{a}l\"{a}inen,
  Alvarez, Bosco, Mancuso, Panza, Nacmias, Boss{\`{u}}, Hanon, Piccardi,
  Annoni, Seripa, Galimberti, Licastro, Soininen, Dartigues, Kamboh,
  Broeckhoven, Lambert, Amouyel, and Campion]{Genin2011}
Genin, E; Hannequin, D; Wallon, D; Sleegers, K; Hiltunen, M; Combarros, O;
  Bullido, M~J; Engelborghs, S; Deyn, P~De; Berr, C; Pasquier, F; Dubois, B;
  Tognoni, G; Fi{\'{e}}vet, N; Brouwers, N; Bettens, K; Arosio, B; Coto, E;
  Zompo, M~Del; Mateo, I; Epelbaum, J; Frank-Garcia, A; Helisalmi, S;
  Porcellini, E; Pilotto, A; Forti, P; Ferri, R; Scarpini, E; Siciliano, G;
  Solfrizzi, V; Sorbi, S; Spalletta, G; Valdivieso, F; Veps\"{a}l\"{a}inen, S;
  Alvarez, V; Bosco, P; Mancuso, M; Panza, F; Nacmias, B; Boss{\`{u}}, P;
  Hanon, O; Piccardi, P; Annoni, G; Seripa, D; Galimberti, D; Licastro, F;
  Soininen, H; Dartigues, J-F; Kamboh, M~I; Broeckhoven, C~Van; Lambert, J~C;
  Amouyel, P, and Campion, D.
\newblock {APOE} and alzheimer disease: a major gene with semi-dominant
  inheritance.
\newblock \emph{Molecular Psychiatry}, 16\penalty0 (9):\penalty0 903--907,
  2011.
\newblock \doi{10.1038/mp.2011.52}.
\newblock URL \url{https://doi.org/10.1038/mp.2011.52}.

\bibitem[Ghisletta et~al.(2010)Ghisletta, Kennedy, Rodrigue, Lindenberger, and
  Raz]{Ghisletta2010}
Ghisletta, Paolo; Kennedy, Kristen~M.; Rodrigue, Karen~M.; Lindenberger, Ulman,
  and Raz, Naftali.
\newblock Adult age differences and the role of cognitive resources in
  perceptual{\textendash}motor skill acquisition: Application of a multilevel
  negative exponential model.
\newblock \emph{The Journals of Gerontology: Series B}, 65B\penalty0
  (2):\penalty0 163--173, January 2010.
\newblock \doi{10.1093/geronb/gbp126}.
\newblock URL \url{https://doi.org/10.1093/geronb/gbp126}.

\bibitem[Golub et~al.(1979)Golub, Heath, and Wahba]{Golub1979}
Golub, Gene~H.; Heath, Michael, and Wahba, Grace.
\newblock Generalized cross-validation as a method for choosing a good ridge
  parameter.
\newblock \emph{Technometrics}, 21\penalty0 (2):\penalty0 215--223, 1979.
\newblock \doi{10.1080/00401706.1979.10489751}.
\newblock URL \url{https://doi.org/10.1080/00401706.1979.10489751}.

\bibitem[Gorbach et~al.(2017)Gorbach, Pudas, Lundquist, Or\"{a}dd, Josefsson,
  Salami, de~Luna, and Nyberg]{Gorbach2017}
Gorbach, Tetiana; Pudas, Sara; Lundquist, Anders; Or\"{a}dd, Greger; Josefsson,
  Maria; Salami, Alireza; de~Luna, Xavier, and Nyberg, Lars.
\newblock Longitudinal association between hippocampus atrophy and
  episodic-memory decline.
\newblock \emph{Neurobiology of Aging}, 51:\penalty0 167--176, March 2017.
\newblock \doi{10.1016/j.neurobiolaging.2016.12.002}.
\newblock URL \url{https://doi.org/10.1016/j.neurobiolaging.2016.12.002}.

\bibitem[Grimm et~al.(2013)Grimm, Zhang, Hamagami, and Mazzocco]{Grimm2013}
Grimm, Kevin; Zhang, Zhiyong; Hamagami, Fumiaki, and Mazzocco, Mich{\`{e}}le.
\newblock Modeling nonlinear change via latent change and latent acceleration
  frameworks: Examining velocity and acceleration of growth trajectories.
\newblock \emph{Multivariate Behavioral Research}, 48\penalty0 (1):\penalty0
  117--143, January 2013.
\newblock \doi{10.1080/00273171.2012.755111}.
\newblock URL \url{https://doi.org/10.1080/00273171.2012.755111}.

\bibitem[Grimm and Ram(2009)]{Grimm2009}
Grimm, Kevin~J. and Ram, Nilam.
\newblock Nonlinear growth models in mplusand {SAS}.
\newblock \emph{Structural Equation Modeling: A Multidisciplinary Journal},
  16\penalty0 (4):\penalty0 676--701, October 2009.
\newblock \doi{10.1080/10705510903206055}.
\newblock URL \url{https://doi.org/10.1080/10705510903206055}.

\bibitem[Gu and Ma(2005)]{Gu2005}
Gu, Chong and Ma, Ping.
\newblock Generalized nonparametric mixed-effect models: Computation and
  smoothing parameter selection.
\newblock \emph{Journal of Computational and Graphical Statistics}, 14\penalty0
  (2):\penalty0 485--504, 2005.
\newblock \doi{10.1198/106186005x47651}.
\newblock URL \url{https://doi.org/10.1198/106186005x47651}.

\bibitem[Hastie and Tibshirani(1986)]{Hastie1986}
Hastie, Trevor and Tibshirani, Robert.
\newblock Generalized additive models.
\newblock \emph{Statistical Science}, 1\penalty0 (3):\penalty0 297--310, 08
  1986.
\newblock \doi{10.1214/ss/1177013604}.
\newblock URL \url{https://doi.org/10.1214/ss/1177013604}.

\bibitem[Hastie and Tibshirani(1993)]{Hastie1993}
Hastie, Trevor and Tibshirani, Robert.
\newblock Varying-coefficient models.
\newblock \emph{Journal of the Royal Statistical Society: Series B
  (Methodological)}, 55\penalty0 (4):\penalty0 757--779, 1993.
\newblock \doi{10.1111/j.2517-6161.1993.tb01939.x}.
\newblock URL \url{https://doi.org/10.1111/j.2517-6161.1993.tb01939.x}.

\bibitem[Hastie et~al.(2008)Hastie, Tibshirani, and Friedman]{Hastie2008}
Hastie, Trevor; Tibshirani, Robert, and Friedman, Jerome.
\newblock Model assessment and selection.
\newblock In \emph{The Elements of Statistical Learning}, pages 219--259.
  Springer New York, December 2008.
\newblock \doi{10.1007/978-0-387-84858-7_7}.
\newblock URL \url{https://doi.org/10.1007/978-0-387-84858-7_7}.

\bibitem[Hoffman(2007)]{Hoffman2007}
Hoffman, Lesa.
\newblock Multilevel models for examining individual differences in
  within-person variation and covariation over time.
\newblock \emph{Multivariate Behavioral Research}, 42\penalty0 (4):\penalty0
  609--629, 2007.
\newblock \doi{10.1080/00273170701710072}.
\newblock URL \url{https://doi.org/10.1080/00273170701710072}.

\bibitem[Hoffman and Stawski(2009)]{Hoffman2009}
Hoffman, Lesa and Stawski, Robert~S.
\newblock Persons as contexts: Evaluating between-person and within-person
  effects in longitudinal analysis.
\newblock \emph{Research in Human Development}, 6\penalty0 (2-3):\penalty0
  97--120, 2009.
\newblock \doi{10.1080/15427600902911189}.
\newblock URL \url{https://doi.org/10.1080/15427600902911189}.

\bibitem[Horn and Donaldson(1976)]{Horn1976}
Horn, John~L. and Donaldson, Gary.
\newblock On the myth of intellectual decline in adulthood.
\newblock \emph{American Psychologist}, 31\penalty0 (10):\penalty0 701--719,
  1976.
\newblock \doi{10.1037/0003-066x.31.10.701}.
\newblock URL \url{https://doi.org/10.1037/0003-066x.31.10.701}.

\bibitem[Ke and Wang(2001)]{Ke2001}
Ke, Chunlei and Wang, Yuedong.
\newblock Semiparametric nonlinear mixed-effects models and their applications.
\newblock \emph{Journal of the American Statistical Association}, 96\penalty0
  (456):\penalty0 1272--1298, 2001.
\newblock \doi{10.1198/016214501753381913}.
\newblock URL \url{https://doi.org/10.1198/016214501753381913}.

\bibitem[Kimeldorf and Wahba(1970)]{Kimeldorf1970}
Kimeldorf, George~S. and Wahba, Grace.
\newblock A correspondence between {Bayesian} estimation on stochastic
  processes and smoothing by splines.
\newblock \emph{The Annals of Mathematical Statistics}, 41\penalty0
  (2):\penalty0 495--502, 1970.
\newblock ISSN 00034851.
\newblock URL \url{http://www.jstor.org/stable/2239347}.

\bibitem[Laird and Ware(1982)]{Laird1982}
Laird, Nan~M. and Ware, James~H.
\newblock Random-effects models for longitudinal data.
\newblock \emph{Biometrics}, 38\penalty0 (4):\penalty0 963--974, 1982.
\newblock ISSN 0006341X, 15410420.
\newblock URL \url{http://www.jstor.org/stable/2529876}.

\bibitem[Lambert et~al.(2001)Lambert, Abrams, Jones, Halligan, and
  Shennan]{Lambert2001}
Lambert, Paul~C.; Abrams, Keith~R.; Jones, David~R.; Halligan, Aidan W.~F., and
  Shennan, Andrew.
\newblock Analysis of ambulatory blood pressure monitor data using a
  hierarchical model incorporating restricted cubic splines and heterogeneous
  within-subject variances.
\newblock \emph{Statistics in Medicine}, 20\penalty0 (24):\penalty0 3789--3805,
  2001.
\newblock \doi{10.1002/sim.1172}.
\newblock URL \url{https://doi.org/10.1002/sim.1172}.

\bibitem[Lin and Zhang(1999)]{Lin1999}
Lin, X. and Zhang, D.
\newblock Inference in generalized additive mixed models by using smoothing
  splines.
\newblock \emph{Journal of the Royal Statistical Society: Series B (Statistical
  Methodology)}, 61\penalty0 (2):\penalty0 381--400, 1999.
\newblock \doi{10.1111/1467-9868.00183}.
\newblock URL \url{https://doi.org/10.1111/1467-9868.00183}.

\bibitem[Lindenberger and P\"{o}tter(1998)]{Lindenberger1998}
Lindenberger, Ulman and P\"{o}tter, Ulrich.
\newblock The complex nature of unique and shared effects in hierarchical
  linear regression: Implications for developmental psychology.
\newblock \emph{Psychological Methods}, 3\penalty0 (2):\penalty0 218--230,
  1998.
\newblock \doi{10.1037/1082-989x.3.2.218}.
\newblock URL \url{https://doi.org/10.1037/1082-989x.3.2.218}.

\bibitem[Lindenberger et~al.(2011)Lindenberger, von Oertzen, Ghisletta, and
  Hertzog]{Lindenberger2011}
Lindenberger, Ulman; von Oertzen, Timo; Ghisletta, Paolo, and Hertzog,
  Christopher.
\newblock Cross-sectional age variance extraction: What's change got to do with
  it?
\newblock \emph{Psychology and Aging}, 26\penalty0 (1):\penalty0 34--47, 2011.
\newblock \doi{10.1037/a0020525}.
\newblock URL \url{https://doi.org/10.1037/a0020525}.

\bibitem[Lindstrom and Bates(1990)]{Lindstrom1990}
Lindstrom, Mary~J. and Bates, Douglas~M.
\newblock Nonlinear mixed effects models for repeated measures data.
\newblock \emph{Biometrics}, 46\penalty0 (3):\penalty0 673, September 1990.
\newblock \doi{10.2307/2532087}.
\newblock URL \url{https://doi.org/10.2307/2532087}.

\bibitem[Marra and Wood(2012)]{Marra2012}
Marra, Giamperio and Wood, Simon~N.
\newblock Coverage properties of confidence intervals for generalized additive
  model components.
\newblock \emph{Scandinavian Journal of Statistics}, 39\penalty0 (1):\penalty0
  53--74, 2012.
\newblock \doi{10.1111/j.1467-9469.2011.00760.x}.
\newblock URL
  \url{https://onlinelibrary.wiley.com/doi/abs/10.1111/j.1467-9469.2011.00760.x}.

\bibitem[McArdle and Epstein(1987)]{McArdle1987}
McArdle, J.~J. and Epstein, David.
\newblock Latent growth curves within developmental structural equation models.
\newblock \emph{Child Development}, 58\penalty0 (1):\penalty0 110, February
  1987.
\newblock \doi{10.2307/1130295}.
\newblock URL \url{https://doi.org/10.2307/1130295}.

\bibitem[Mehta and West(2000)]{Mehta2000}
Mehta, Paras~D. and West, Stephen~G.
\newblock Putting the individual back into individual growth curves.
\newblock \emph{Psychological Methods}, 5\penalty0 (1):\penalty0 23--43, 2000.
\newblock \doi{10.1037/1082-989x.5.1.23}.
\newblock URL \url{https://doi.org/10.1037/1082-989x.5.1.23}.

\bibitem[Meredith and Kruschke(2019)]{Meredith2019}
Meredith, Mike and Kruschke, John.
\newblock \emph{HDInterval: Highest (Posterior) Density Intervals}, 2019.
\newblock URL \url{https://CRAN.R-project.org/package=HDInterval}.
\newblock R package version 0.2.2.

\bibitem[Meredith and Tisak(1990)]{Meredith1990}
Meredith, William and Tisak, John.
\newblock Latent curve analysis.
\newblock \emph{Psychometrika}, 55\penalty0 (1):\penalty0 107--122, 1990.
\newblock \doi{10.1007/bf02294746}.
\newblock URL \url{https://doi.org/10.1007/bf02294746}.

\bibitem[Molenaar(2004)]{Molenaar2004}
Molenaar, Peter C.~M.
\newblock A manifesto on psychology as idiographic science: Bringing the person
  back into scientific psychology, this time forever.
\newblock \emph{Measurement: Interdisciplinary Research {\&} Perspective},
  2\penalty0 (4):\penalty0 201--218, 2004.
\newblock \doi{10.1207/s15366359mea0204_1}.
\newblock URL \url{https://doi.org/10.1207/s15366359mea0204_1}.

\bibitem[Molenberghs and Fitzmaurice(2009)]{Molenberghs2009}
Molenberghs, G. and Fitzmaurice, G.
\newblock Incomplete data: Introduction and overview.
\newblock In Fitzmaurice, G.; Davidian, M.; Verbeke, G., and Molenberghs, G.,
  editors, \emph{Longitudinal Data Analysis}, pages 395--408. Chapman and Hall,
  London, U.K., 2009.

\bibitem[Morrell et~al.(2009)Morrell, Brant, and Ferrucci]{Morrell2009}
Morrell, C.~H.; Brant, L.~J., and Ferrucci, L.
\newblock Model choice can obscure results in longitudinal studies.
\newblock \emph{The Journals of Gerontology Series A: Biological Sciences and
  Medical Sciences}, 64A\penalty0 (2):\penalty0 215--222, 2009.
\newblock \doi{10.1093/gerona/gln024}.
\newblock URL \url{https://doi.org/10.1093/gerona/gln024}.

\bibitem[Newsom(2015)]{Newsom2015}
Newsom, Jason~T.
\newblock \emph{Longitudinal Structural Equation Modeling}.
\newblock Routledge, 2015.

\bibitem[Nychka(1988)]{Nychka1988}
Nychka, Douglas.
\newblock Bayesian confidence intervals for smoothing splines.
\newblock \emph{Journal of the American Statistical Association}, 83\penalty0
  (404):\penalty0 1134--1143, 1988.
\newblock \doi{10.1080/01621459.1988.10478711}.
\newblock URL \url{https://doi.org/10.1080/01621459.1988.10478711}.

\bibitem[Oud and Jansen(2000)]{Oud2000}
Oud, Johan H.~L. and Jansen, Robert A. R.~G.
\newblock Continuous time state space modeling of panel data by means of {SEM}.
\newblock \emph{Psychometrika}, 65\penalty0 (2):\penalty0 199--215, 2000.
\newblock \doi{10.1007/bf02294374}.
\newblock URL \url{https://doi.org/10.1007/bf02294374}.

\bibitem[Pedersen et~al.(2019)Pedersen, Miller, Simpson, and
  Ross]{Pedersen2019}
Pedersen, Eric~J.; Miller, David~L.; Simpson, Gavin~L., and Ross, Noam.
\newblock Hierarchical generalized additive models in ecology: an introduction
  with mgcv.
\newblock \emph{{PeerJ}}, 7:\penalty0 e6876, 2019.
\newblock \doi{10.7717/peerj.6876}.
\newblock URL \url{https://doi.org/10.7717/peerj.6876}.

\bibitem[{R Core Team}(2019)]{Rcore}
{R Core Team}, .
\newblock \emph{R: A Language and Environment for Statistical Computing}.
\newblock R Foundation for Statistical Computing, Vienna, Austria, 2019.
\newblock URL \url{https://www.R-project.org/}.

\bibitem[Ram and Grimm(2007)]{Ram2007}
Ram, Nilam and Grimm, Kevin.
\newblock Using simple and complex growth models to articulate developmental
  change: Matching theory to method.
\newblock \emph{International Journal of Behavioral Development}, 31\penalty0
  (4):\penalty0 303--316, July 2007.
\newblock \doi{10.1177/0165025407077751}.
\newblock URL \url{https://doi.org/10.1177/0165025407077751}.

\bibitem[Raz et~al.(2005)Raz, Lindenberger, Rodrigue, Kennedy, Head,
  Williamson, Dahle, Gerstorf, and Acker]{Raz2005}
Raz, Naftali; Lindenberger, Ulman; Rodrigue, Karen~M.; Kennedy, Kristen~M.;
  Head, Denise; Williamson, Adrienne; Dahle, Cheryl; Gerstorf, Denis, and
  Acker, James~D.
\newblock Regional brain changes in aging healthy adults: General trends,
  individual differences and modifiers.
\newblock \emph{Cerebral Cortex}, 15\penalty0 (11):\penalty0 1676--1689, 2005.
\newblock \doi{10.1093/cercor/bhi044}.
\newblock URL \url{https://doi.org/10.1093/cercor/bhi044}.

\bibitem[Raz et~al.(2010)Raz, Ghisletta, Rodrigue, Kennedy, and
  Lindenberger]{Raz2010}
Raz, Naftali; Ghisletta, Paolo; Rodrigue, Karen~M.; Kennedy, Kristen~M., and
  Lindenberger, Ulman.
\newblock Trajectories of brain aging in middle-aged and older adults: Regional
  and individual differences.
\newblock \emph{{NeuroImage}}, 51\penalty0 (2):\penalty0 501--511, 2010.
\newblock \doi{10.1016/j.neuroimage.2010.03.020}.
\newblock URL \url{https://doi.org/10.1016/j.neuroimage.2010.03.020}.

\bibitem[Reuter et~al.(2012)Reuter, Schmansky, Rosas, and Fischl]{Reuter2012}
Reuter, Martin; Schmansky, Nicholas~J.; Rosas, H.~Diana, and Fischl, Bruce.
\newblock Within-subject template estimation for unbiased longitudinal image
  analysis.
\newblock \emph{{NeuroImage}}, 61\penalty0 (4):\penalty0 1402--1418, 2012.
\newblock \doi{10.1016/j.neuroimage.2012.02.084}.
\newblock URL \url{https://doi.org/10.1016/j.neuroimage.2012.02.084}.

\bibitem[Ruppert et~al.(2003)Ruppert, Wand, and Carroll]{Ruppert2003}
Ruppert, David; Wand, M.~P., and Carroll, R.~J.
\newblock \emph{Semiparametric Regression}.
\newblock Cambridge University Press, Cambridge, U.K., 2003.

\bibitem[Salthouse(2013)]{Salthouse2013}
Salthouse, Timothy~A.
\newblock Within-cohort age-related differences in cognitive functioning.
\newblock \emph{Psychological Science}, 24\penalty0 (2):\penalty0 123--130,
  January 2013.
\newblock \doi{10.1177/0956797612450893}.
\newblock URL \url{https://doi.org/10.1177/0956797612450893}.

\bibitem[Salthouse(2014)]{Salthouse2014}
Salthouse, Timothy~A.
\newblock Why are there different age relations in cross-sectional and
  longitudinal comparisons of cognitive functioning?
\newblock \emph{Current Directions in Psychological Science}, 23\penalty0
  (4):\penalty0 252--256, August 2014.
\newblock \doi{10.1177/0963721414535212}.
\newblock URL \url{https://doi.org/10.1177/0963721414535212}.

\bibitem[Salthouse(2019)]{Salthouse2019}
Salthouse, Timothy~A.
\newblock Trajectories of normal cognitive aging.
\newblock \emph{Psychology and Aging}, 34\penalty0 (1):\penalty0 17--24,
  February 2019.
\newblock \doi{10.1037/pag0000288}.
\newblock URL \url{https://doi.org/10.1037/pag0000288}.

\bibitem[Schaie et~al.(1973)Schaie, Labouvie, and Buech]{Schaie1973}
Schaie, K.~Warner; Labouvie, Gisela~V., and Buech, Barbara~U.
\newblock Generational and cohort-specific differences in adult cognitive
  functioning: A fourteen-year study of independent samples.
\newblock \emph{Developmental Psychology}, 9\penalty0 (2):\penalty0 151--166,
  1973.
\newblock \doi{10.1037/h0035093}.
\newblock URL \url{https://doi.org/10.1037/h0035093}.

\bibitem[Simpson(2016)]{Simpson2016}
Simpson, Gavin.
\newblock \emph{Simultaneous intervals for smooths revisited}, 2016.
\newblock
  https://fromthebottomoftheheap.net/2016/12/15/simultaneous-interval-revisited/
  (accessed July 10, 2020).

\bibitem[Sliwinski et~al.(2010)Sliwinski, Hoffman, and Hofer]{Sliwinski2010}
Sliwinski, Martin; Hoffman, Lesa, and Hofer, Scott~M.
\newblock Evaluating convergence of within-person change and between-person age
  differences in age-heterogeneous longitudinal studies.
\newblock \emph{Research in Human Development}, 7\penalty0 (1):\penalty0
  45--60, 2010.
\newblock \doi{10.1080/15427600903578169}.
\newblock URL \url{https://doi.org/10.1080/15427600903578169}.

\bibitem[S{\o}rensen et~al.(2021)S{\o}rensen, Brandmaier, Maci{\`{a}}, Ebmeier,
  Ghisletta, Kievit, Mowinckel, Walhovd, Westerhausen, and Fjell]{Sorensen2021}
S{\o}rensen, {\O}ystein; Brandmaier, Andreas~M.; Maci{\`{a}}, D{\'{\i}}dac;
  Ebmeier, Klaus; Ghisletta, Paolo; Kievit, Rogier~A.; Mowinckel, Athanasia~M.;
  Walhovd, Kristine~B.; Westerhausen, Rene, and Fjell, Anders.
\newblock Meta-analysis of generalized additive models in neuroimaging studies.
\newblock \emph{{NeuroImage}}, 224:\penalty0 117416, January 2021.
\newblock \doi{10.1016/j.neuroimage.2020.117416}.
\newblock URL \url{https://doi.org/10.1016/j.neuroimage.2020.117416}.

\bibitem[Sullivan et~al.(2015)Sullivan, Shadish, and Steiner]{Sullivan2015}
Sullivan, Kristynn~J.; Shadish, William~R., and Steiner, Peter~M.
\newblock An introduction to modeling longitudinal data with generalized
  additive models: Applications to single-case designs.
\newblock \emph{Psychological Methods}, 20\penalty0 (1):\penalty0 26--42, 2015.
\newblock \doi{10.1037/met0000020}.
\newblock URL \url{https://doi.org/10.1037/met0000020}.

\bibitem[Thompson et~al.(2017)Thompson, Andreassen, Arias-Vasquez, Bearden,
  Boedhoe, Brouwer, Buckner, Buitelaar, Bulayeva, Cannon, Cohen, Conrod, Dale,
  Deary, Dennis, de~Reus, Desrivieres, Dima, Donohoe, Fisher, Fouche, Francks,
  Frangou, Franke, Ganjgahi, Garavan, Glahn, Grabe, Guadalupe, Gutman,
  Hashimoto, Hibar, Holland, Hoogman, Pol, Hosten, Jahanshad, Kelly, Kochunov,
  Kremen, Lee, Mackey, Martin, Mazoyer, McDonald, Medland, Morey, Nichols,
  Paus, Pausova, Schmaal, Schumann, Shen, Sisodiya, Smit, Smoller, Stein,
  Stein, Toro, Turner, van~den Heuvel, van~den Heuvel, van Erp, van Rooij,
  Veltman, Walter, Wang, Wardlaw, Whelan, Wright, and Ye]{Thompson2017}
Thompson, Paul~M.; Andreassen, Ole~A.; Arias-Vasquez, Alejandro; Bearden,
  Carrie~E.; Boedhoe, Premika~S.; Brouwer, Rachel~M.; Buckner, Randy~L.;
  Buitelaar, Jan~K.; Bulayeva, Kazima~B.; Cannon, Dara~M.; Cohen, Ronald~A.;
  Conrod, Patricia~J.; Dale, Anders~M.; Deary, Ian~J.; Dennis, Emily~L.;
  de~Reus, Marcel~A.; Desrivieres, Sylvane; Dima, Danai; Donohoe, Gary; Fisher,
  Simon~E.; Fouche, Jean-Paul; Francks, Clyde; Frangou, Sophia; Franke,
  Barbara; Ganjgahi, Habib; Garavan, Hugh; Glahn, David~C.; Grabe, Hans~J.;
  Guadalupe, Tulio; Gutman, Boris~A.; Hashimoto, Ryota; Hibar, Derrek~P.;
  Holland, Dominic; Hoogman, Martine; Pol, Hilleke E.~Hulshoff; Hosten,
  Norbert; Jahanshad, Neda; Kelly, Sinead; Kochunov, Peter; Kremen, William~S.;
  Lee, Phil~H.; Mackey, Scott; Martin, Nicholas~G.; Mazoyer, Bernard; McDonald,
  Colm; Medland, Sarah~E.; Morey, Rajendra~A.; Nichols, Thomas~E.; Paus, Tomas;
  Pausova, Zdenka; Schmaal, Lianne; Schumann, Gunter; Shen, Li; Sisodiya,
  Sanjay~M.; Smit, Dirk~J.A.; Smoller, Jordan~W.; Stein, Dan~J.; Stein,
  Jason~L.; Toro, Roberto; Turner, Jessica~A.; van~den Heuvel, Martijn~P.;
  van~den Heuvel, Odile~L.; van Erp, Theo~G.M.; van Rooij, Daan; Veltman,
  Dick~J.; Walter, Henrik; Wang, Yalin; Wardlaw, Joanna~M.; Whelan,
  Christopher~D.; Wright, Margaret~J., and Ye, Jieping.
\newblock {ENIGMA} and the individual: Predicting factors that affect the brain
  in 35 countries worldwide.
\newblock \emph{{NeuroImage}}, 145:\penalty0 389--408, 2017.
\newblock \doi{10.1016/j.neuroimage.2015.11.057}.
\newblock URL \url{https://doi.org/10.1016/j.neuroimage.2015.11.057}.

\bibitem[Thompson et~al.(2011)Thompson, Hallmayer, and and]{Thompson2011}
Thompson, Wesley~K.; Hallmayer, Joachim, and and, Ruth~O'Hara.
\newblock Design considerations for characterizing psychiatric trajectories
  across the lifespan: Application to effects of {APOE}-$\epsilon$4 on cerebral
  cortical thickness in {Alzheimer}'s disease.
\newblock \emph{American Journal of Psychiatry}, 168\penalty0 (9):\penalty0
  894--903, September 2011.
\newblock \doi{10.1176/appi.ajp.2011.10111690}.
\newblock URL \url{https://doi.org/10.1176/appi.ajp.2011.10111690}.

\bibitem[Tian et~al.(2015)Tian, Studenski, Resnick, Davatzikos, and
  Ferrucci]{Tian2015}
Tian, Qu; Studenski, Stephanie~A.; Resnick, Susan~M.; Davatzikos, Christos, and
  Ferrucci, Luigi.
\newblock Midlife and late-life cardiorespiratory fitness and brain volume
  changes in late adulthood: Results from the baltimore longitudinal study of
  aging.
\newblock \emph{The Journals of Gerontology Series A: Biological Sciences and
  Medical Sciences}, 71\penalty0 (1):\penalty0 124--130, April 2015.
\newblock \doi{10.1093/gerona/glv041}.
\newblock URL \url{https://doi.org/10.1093/gerona/glv041}.

\bibitem[Walhovd et~al.(2018)Walhovd, Fjell, Westerhausen, Nyberg, Ebmeier,
  Lindenberger, Bartres-Faz, Baare, Siebner, Henson, and et~al.]{Walhovd2018}
Walhovd, K.B.; Fjell, A.M.; Westerhausen, R.; Nyberg, L.; Ebmeier, K.P.;
  Lindenberger, U.; Bartres-Faz, D.; Baare, W.F.C.; Siebner, H.R.; Henson, R.,
  and et~al., .
\newblock Healthy minds 0–100 years: Optimising the use of {E}uropean brain
  imaging cohorts (“{L}ifebrain”).
\newblock \emph{European Psychiatry}, 47:\penalty0 76–77, 2018.
\newblock \doi{10.1016/j.eurpsy.2017.10.005}.

\bibitem[Walhovd et~al.(2016)Walhovd, Krogsrud, Amlien, Bartsch, Bj{\o}rnerud,
  Due-T{\o}nnessen, Grydeland, Hagler, H{\aa}berg, Kremen, Ferschmann, Nyberg,
  Panizzon, Rohani, Skranes, Storsve, S{\o}lsnes, Tamnes, Thompson, Reuter,
  Dale, and Fjell]{Walhovd2016}
Walhovd, Kristine~B.; Krogsrud, Stine~K.; Amlien, Inge~K.; Bartsch, Hauke;
  Bj{\o}rnerud, Atle; Due-T{\o}nnessen, Paulina; Grydeland, H{\aa}kon; Hagler,
  Donald~J.; H{\aa}berg, Asta~K.; Kremen, William~S.; Ferschmann, Lia; Nyberg,
  Lars; Panizzon, Matthew~S.; Rohani, Darius~A.; Skranes, Jon; Storsve,
  Andreas~B.; S{\o}lsnes, Anne~Elisabeth; Tamnes, Christian~K.; Thompson,
  Wesley~K.; Reuter, Chase; Dale, Anders~M., and Fjell, Anders~M.
\newblock Neurodevelopmental origins of lifespan changes in brain and
  cognition.
\newblock \emph{Proceedings of the National Academy of Sciences}, 113\penalty0
  (33):\penalty0 9357--9362, 2016.
\newblock \doi{10.1073/pnas.1524259113}.
\newblock URL \url{https://doi.org/10.1073/pnas.1524259113}.

\bibitem[Walhovd et~al.(2019)Walhovd, Fjell, S{\o}rensen, Mowinckel, Reinbold,
  Idland, Watne, Franke, Dobricic, Kilpert, Bertram, and Wang]{Walhovd2019}
Walhovd, Kristine~B; Fjell, Anders~M.; S{\o}rensen, {\O}ystein; Mowinckel,
  Athanasia~Monica; Reinbold, C{\'{e}}line~Sonja; Idland, Ane-Victoria; Watne,
  Leiv~Otto; Franke, Andre; Dobricic, Valerijia; Kilpert, Fabian; Bertram,
  Lars, and Wang, Yunpeng.
\newblock Genetic risk for alzheimer{\textquoteright}s disease predicts
  hippocampal volume through the lifespan.
\newblock \emph{bioRxiv}, 2019.
\newblock \doi{10.1101/711689}.
\newblock URL \url{https://www.biorxiv.org/content/early/2019/07/23/711689}.

\bibitem[Wickham(2016)]{Wickham2016}
Wickham, Hadley.
\newblock \emph{ggplot2: Elegant Graphics for Data Analysis}.
\newblock Springer-Verlag, New York, New York, USA, 2016.
\newblock ISBN 978-3-319-24277-4.
\newblock URL \url{https://ggplot2.tidyverse.org}.

\bibitem[Wood and Scheipl(2017)]{Wood2017gamm4}
Wood, Simon and Scheipl, Fabian.
\newblock \emph{gamm4: Generalized Additive Mixed Models using 'mgcv' and
  'lme4'}, 2017.
\newblock URL \url{https://CRAN.R-project.org/package=gamm4}.
\newblock R package version 0.2-5.

\bibitem[Wood(2003)]{Wood2003}
Wood, Simon~N.
\newblock Thin plate regression splines.
\newblock \emph{Journal of the Royal Statistical Society: Series B (Statistical
  Methodology)}, 65\penalty0 (1):\penalty0 95--114, 2003.
\newblock \doi{10.1111/1467-9868.00374}.
\newblock URL
  \url{https://rss.onlinelibrary.wiley.com/doi/abs/10.1111/1467-9868.00374}.

\bibitem[Wood(2004)]{Wood2004}
Wood, Simon~N.
\newblock Stable and efficient multiple smoothing parameter estimation for
  generalized additive models.
\newblock \emph{Journal of the American Statistical Association}, 99\penalty0
  (467):\penalty0 673--686, 2004.
\newblock \doi{10.1198/016214504000000980}.
\newblock URL \url{https://doi.org/10.1198/016214504000000980}.

\bibitem[Wood(2006)]{Wood2006}
Wood, Simon~N.
\newblock Low-rank scale-invariant tensor product smooths for generalized
  additive mixed models.
\newblock \emph{Biometrics}, 62\penalty0 (4):\penalty0 1025--1036, 2006.
\newblock \doi{10.1111/j.1541-0420.2006.00574.x}.
\newblock URL
  \url{https://onlinelibrary.wiley.com/doi/abs/10.1111/j.1541-0420.2006.00574.x}.

\bibitem[Wood(2010)]{Wood2010}
Wood, Simon~N.
\newblock Fast stable restricted maximum likelihood and marginal likelihood
  estimation of semiparametric generalized linear models.
\newblock \emph{Journal of the Royal Statistical Society: Series B (Statistical
  Methodology)}, 73\penalty0 (1):\penalty0 3--36, 2010.
\newblock \doi{10.1111/j.1467-9868.2010.00749.x}.
\newblock URL \url{https://doi.org/10.1111/j.1467-9868.2010.00749.x}.

\bibitem[Wood et~al.(2012)Wood, Scheipl, and Faraway]{Wood2012b}
Wood, Simon~N.; Scheipl, Fabian, and Faraway, Julian~J.
\newblock Straightforward intermediate rank tensor product smoothing in mixed
  models.
\newblock \emph{Statistics and Computing}, 23\penalty0 (3):\penalty0 341--360,
  2012.
\newblock \doi{10.1007/s11222-012-9314-z}.
\newblock URL \url{https://doi.org/10.1007/s11222-012-9314-z}.

\bibitem[Wood(2017)]{Wood2017}
Wood, S.N.
\newblock \emph{Generalized Additive Models: An Introduction with R}.
\newblock Chapman and Hall/CRC, Boca Raton, Florida, USA, 2nd edition, 2017.

\bibitem[Zeger and Liang(1992)]{Zeger1992}
Zeger, Scott~L. and Liang, Kung-Yee.
\newblock An overview of methods for the analysis of longitudinal data.
\newblock \emph{Statistics in Medicine}, 11\penalty0 (14-15):\penalty0
  1825--1839, 1992.
\newblock \doi{10.1002/sim.4780111406}.
\newblock URL \url{https://doi.org/10.1002/sim.4780111406}.

\end{thebibliography}
